\documentclass[review]{elsarticle}
\makeatletter
\def\ps@pprintTitle{%
 \let\@oddhead\@empty
 \let\@evenhead\@empty
 \def\@oddfoot{\centerline{\thepage}}%
 \let\@evenfoot\@oddfoot}
\makeatother

\usepackage{lineno,hyperref}

\usepackage{multirow}
\usepackage{placeins}
\usepackage{graphicx}
\usepackage{subcaption}
\usepackage{comment}
\usepackage{xcolor}
\DeclareGraphicsExtensions{.pdf,.jpeg,.png,.jpg}
\modulolinenumbers[5]

\journal{Nuclear Instruments and Methods in Physics Research Section A}

\begin{document}

\begin{frontmatter}

\title{A simulational study of the indirect geometry neutron spectrometer, BIFROST at the European Spallation Source, from neutron source position to detector position} 

\author[mymainaddress,mysecondaryaddress,mytertiaryaddress]{M.~Klausz\corref{mycorrespondingauthor}}
\cortext[mycorrespondingauthor]{Corresponding author}
\ead{milan.klausz@energia.mta.hu}
\author[mysecondaryaddress]{K.~Kanaki}
\author[mysecondaryaddress]{T.~Kittelmann}
\author[mysecondaryaddress,myquinaryaddress]{R.~Toft-Petersen}
\author[myquaternaryaddress]{J.O.~Birk}
\author[myquaternaryaddress]{M.A.~Olsen}
\author[mymainaddress,mytertiaryaddress]{P.~Zagyvai}
\author[mysecondaryaddress,mysenaryaddress]{R.J.~Hall-Wilton}

\address[mymainaddress]{Hungarian Academy of Sciences, Centre for Energy Research, 1525 Budapest 114., P.O. Box 49., Hungary}
\address[mysecondaryaddress]{European Spallation Source ESS ERIC, P.O Box 176, SE-221 00 Lund, Sweden}
\address[mytertiaryaddress]{Budapest University of Technology and Economics, Institute of Nuclear Techniques, 1111 Budapest, M\H uegyetem rakpart 9., Hungary}
\address[myquinaryaddress]{Technical University of Denmark, Department of Physics, DK-2800 Kongens Lyngby, Denmark}
\address[myquaternaryaddress]{Nanoscience Center, Niels Bohr Institute, University of Copenhagen, DK-2100 Copenhagen {\O}, Denmark}
\address[mysenaryaddress]{Universit\`a degli Studi di Milano-Bicocca, Piazza della Scienza 3, 20126 Milano, Italy}

\begin{abstract}

The European Spallation Source (ESS) is intended to become the most powerful spallation neutron source in the world and the flagship of neutron science in the upcoming decades. The exceptionally high neutron flux will provide unique opportunities for scientific experiments, but also set high requirements for the detectors.
One of the most challenging aspects is the rate capability and in particular the peak instantaneous rate capability, i.e.\,the number of neutrons hitting the detector per channel or cm$^2$ at the peak of the neutron pulse. 
 
The primary purpose of this paper is to estimate the incident rates that are anticipated for the BIFROST instrument planned for ESS, and also to demonstrate the use of powerful simulation tools for the correct interpretation of neutron transport in crystalline materials. 

A full simulation model of the instrument from source to detector position, implemented with the use of multiple simulation software packages is presented.
For a single detector tube instantaneous incident rates with a maximum of 1.7~GHz for a Bragg peak from a single crystal, and 0.3~MHz for a vanadium sample are found. 

This paper also includes the first application of a new pyrolytic graphite model, and a comparison of different simulation tools to highlight their strengths and weaknesses.

\end{abstract}


\end{frontmatter}


\section{Introduction}

The European Spallation Source (ESS) ERIC~\cite{esstdr, ess2018, ANDERSEN2020163402} is designed to operate using the most powerful spallation neutron source in the world, and to provide unprecedentedly high neutron fluxes for instruments of various neutron techniques. One of these instruments is BIFROST~\cite{bifrost_proposal, Freeman:2014msa}, a high flux, indirect geometry, cold spectrometer, optimised for small samples and extreme environments. 
BIFROST is primarily intended for single-crystal inelastic scattering studies, providing exceptional flux as it can operate in a white beam mode.
This flux allows for entirely new options for detailed investigations of complex multimode dynamics, hybrid modes, electro-magnons, spin wave continua and gap studies, under extreme conditions with controlled temperature, pressure, and magnetic fields.

Harnessing the full ESS pulse by employing a polychromatic beam carries a high potential risk for detector rates that can saturate the detectors and therefore degrade the performance of the instrument. 
The chosen detector technology of BIFROST, position sensitive $^3$He tubes are the ``gold standard'' for neutron detection \cite{Knoll}. They are however quite rate limited. Non-position sensitive tubes saturate at 100~kHz; however for position sensitive $^3$He tubes, operation at instantaneous rate above 30~kHz can be problematic. 
The exact rate capability of a detector is dependent on the details of readout electronics. It is therefore essential to evaluate the rates anticipated for high-flux instruments~\cite{ratesPaper, Stefanescu_2017, MB2017}, in order to extract the respective detector requirements.

Monte Carlo simulation plays a key role in the development and characterisation of instruments as a reliable, cheap and versatile tool~\cite{KANAKI2018386}. 
Feedback from simulations taken into account in the development of the instrument design
can reduce the number of physical prototypes needed, and also enables the quantification of otherwise unmeasurable properties. 
This is particularly the case for complicated instruments, such as BIFROST. 
Development of complete and detailed instrument simulation models enables simulations from source to detectors, offering the opportunity to discover and decouple otherwise undetectable cumulative effects. 
These models can provide valuable input for developing calibration and correction routines for data reduction and analysis, and could later be used for experiment planning by users, to predict experimental conditions from specific proposed samples, i.e.\,sample size and composition.

To make detailed full instrument simulations possible, advanced simulation tools have been developed such as NCrystal~\cite{KANAKI2018386, CAI2020106851}, which enables Monte Carlo simulations of thermal neutrons in crystals, and Monte Carlo Particle List (MCPL)~\cite{mcplpaper,mcplgithub}, which enables communication between different software packages.
These tools can greatly enhance the capabilities of the existing and widely used simulation software such as McStas~\cite{mcstas1, mcstas2} and Geant4~\cite{geant4a,geant4b,geant4c_inpresscorrectedproof}, and enable implementation of full simulation models by connecting them to do a chain of simulations, using each of them where they are the most capable.
Such simulation of full instruments is a novel method that has been applied only in a handful of cases~\cite{ratesPaper, Stefanescu_2019}.
 
In this study multiple Monte Carlo simulation tools are used together to implement a full simulation model of the BIFROST instrument from the neutron source to the detector position, featuring the first application the new NCrystal pyrolytic graphite material model.
This full instrument model is used to estimate the incident detector rates that are anticipated in the case of the highest possible incident neutron intensity -- that will be consequently mentioned as 'highest-case' -- and normal-use scenarios, intended to serve as basis for the determination of detector requirements for rate capability. 
For a part of the instrument both a McStas and a Geant4 model are implemented, facilitating the cross-validation of results and the comparison of the two simulation software.

In the following sections, the instrument and simulation models geometries and tools are introduced first, followed by the presentation of incident rates for elastic peaks for various instrument parameters, sample types, sizes and mosaicities, along with a demonstration of the differences between McStas and Geant4 simulation results. The study concludes with the demonstration of the elastic signal of a standard calibration sample.

 \FloatBarrier
\section{BIFROST instrument and simulation model} 

\subsection{The BIFROST instrument \label{subsec:instrument}}

BIFROST is a 162~m long cold neutron spectrometer intended to be built as a first tranche instrument for ESS.  
It combines an indirect geometry time-of-flight (ToF) front end, and an angular and energy multiplexed crystal analyser-based back end.
A back end similar in principle to that installed recently at the CAMEA spectrometer at the PSI~\cite{camea_prototype, camea_groitl}.
BIFROST is designed~\cite{Holm_Dahlin_2019} to maximise the use of the ESS long pulse to enable measurements on small samples and study dynamic properties, 
transporting a 1.7~{\AA} wavelength band to the sample and investigating an energy transfer range from --3 to +55~meV.
The envisioned application fields include materials science, magnetism, life sciences and planetary sciences~\cite{Freeman:2014msa}.

\begin{figure}[!h]  
  \centering
  \begin{subfigure}{\textwidth} 
      \includegraphics[width=\textwidth]{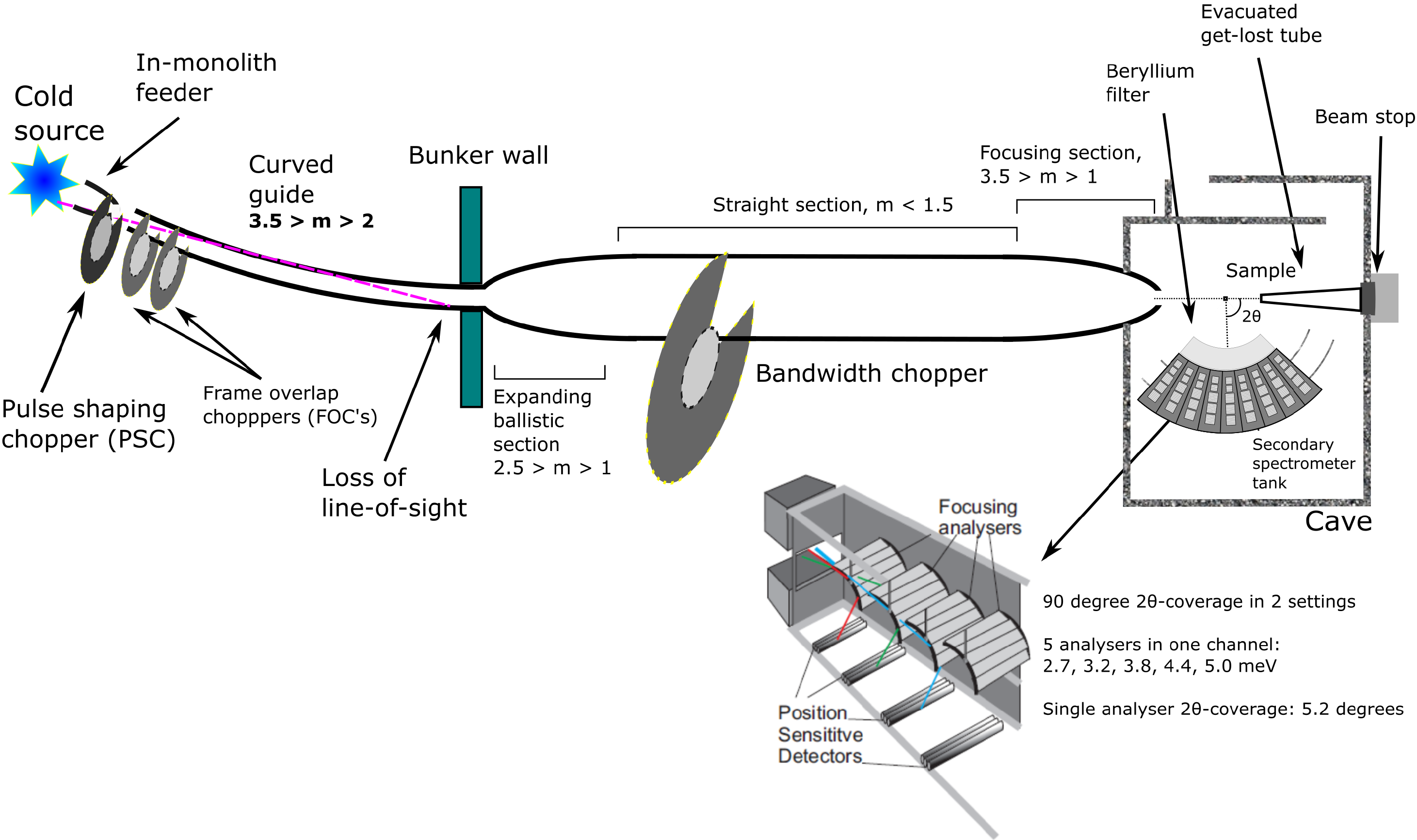}
     \end{subfigure}
 \caption{\footnotesize Schematic model of BIFROST from source to detection position. From~\cite{essbifrost}.}
 \label{fig:bifrost_schematic}
\end{figure}

The instrument consists of three main technical subsystems: the beam transport and conditioning system, the sample exposure system and the scattering characterisation system. The schematic model of the instrument is depicted in Fig.~\ref{fig:bifrost_schematic}.

The beam transport and conditioning system is relatively simple. 
It has a curved guide section inside the bunker to lose line-of-sight, and four choppers as the only moving parts.
Three of these choppers are placed inside the bunker and the fourth one is placed in the middle of the instrument.
The first one is a pulse-shaping chopper, that is the only chopper determining the energy resolution. The other three -- two Frame Overlap and one Bandwidth Chopper -- serve to sort out unwanted frames from the fast pulse-shaping chopper and to avoid pulse overlap at the sample position respectively.
The pulse-shaping chopper can reduce the ESS pulse width by a factor of up to 30 to match the best analyser resolution, or allow the full pulse to reach the sample that will result in a relaxed resolution but an order of 10$^{10}$~n/s/cm$^2$ flux on sample. 
It is this mode which poses the greatest rate challenge for the detectors.

The sample exposure system allows measurements with strong magnetic field, high pressure and cryogenic temperatures.
One of the main limitations today in single crystal neutron spectroscopy is that measurements are only possible with large samples, which are not available for many sample types, but BIFROST will enable the study of sub-cubic millimetre samples thanks to its exceptional flux on sample and the efficient scattering characterisation system.

\begin{figure}[!h]  
  \centering
  \begin{subfigure}{0.35\textwidth} 
      \includegraphics[width=\textwidth]{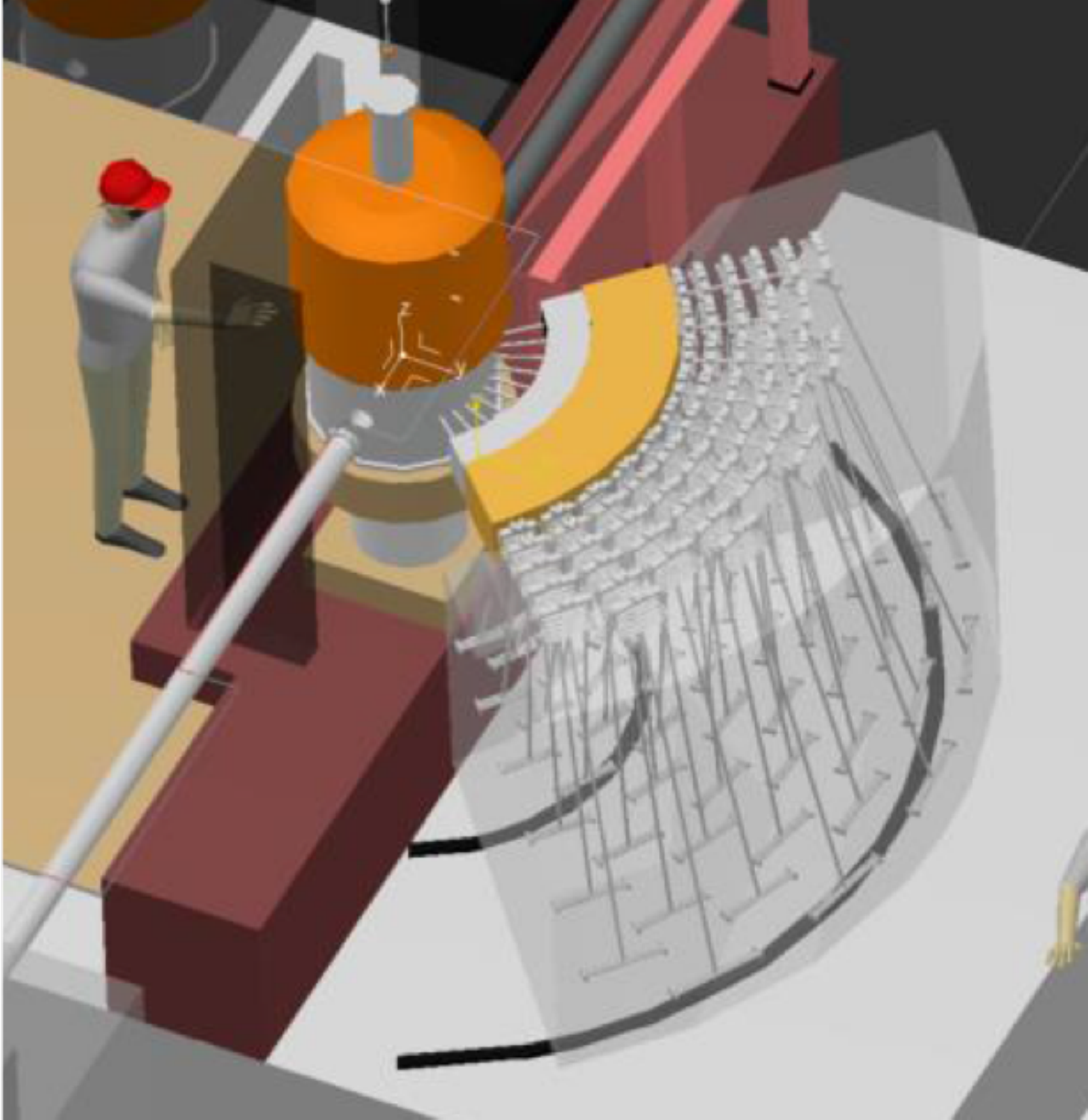}
      \caption{}
     \end{subfigure}
      \begin{subfigure}{0.64\textwidth} 
      \includegraphics[width=\textwidth]{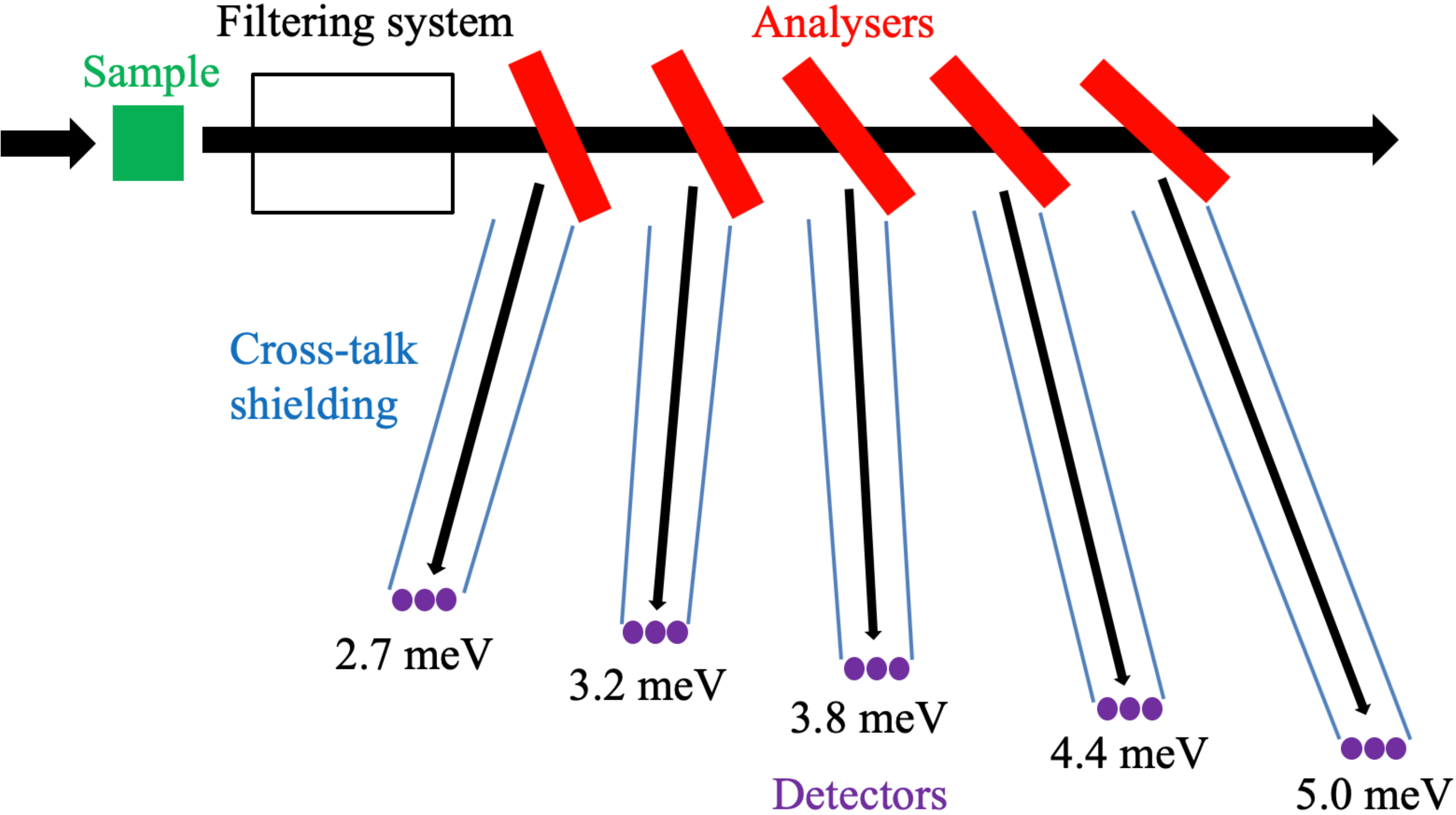}
      \caption{}
      \label{fig:scsSchematic}
     \end{subfigure}
 \caption{\footnotesize The BIFROST scattering characterisation system. 3D model of the secondary spectrometer tank from~\cite{bifrost_proposal} (a) and side view sketch of the sample and a single Q-channel with five energy channels in it (b).}
  \label{fig:scs}
\end{figure}

The scattering characterisation system in Fig.~\ref{fig:scs} consists of the filtering system and the secondary spectrometer tank, covering a 90$^{\circ}$ scattering angle in the horizontal plane in two tank settings, in the 7$^{\circ}$--135$^{\circ}$ 2$\Theta$ range. 
The filtering system, that is essential for background reduction on BIFROST, includes a cooled beryllium filter with roughly 90\% transmission of neutrons with an energy below 5~meV(4.05~\AA) but very low transmission of neutrons with energies above, and coated lamellas as a radial collimator~\cite{GROITL201699}.

The secondary spectrometer tank houses multiple sets of analysers and detectors for different neutron energies, arranged in nine so called Q-channels. 
Depending on the scattering angle of a neutron on the sample, it enters one of the Q-channels, which are separated by cross-talk shielding between them.
In each Q-channel, several crystal analyser arcs are placed one behind the other to select different final energies by scattering neutrons vertically (down) towards the corresponding set of position sensitive detectors, employing Rowland focusing (see Fig.~\ref{fig:scsSchematic}). Further cross-talk shielding is applied to ensure that neutrons can reach the detectors only by scattering from the corresponding analyser arcs.

With this arrangement, BIFROST utilises a variant of a novel analyser setup called CAMEA~\cite{camea_groitl, camea_prototype}, an acronym for Continuous Angle Multiple Energy Analysis.
Enabling multi-energy analysis in a single Q-channel by placing the analysers for higher neutron energies behind the ones for lower energies is possible due to the high transparency of the 1~mm thin highly-oriented pyrolytic graphite blades~\cite{Mildner:hw0086}.
The blades to be used have high mosaicity (60~arcmin)  
to apply the prismatic analyser concept~\cite{prismaticAnalyserConcept}, using $^3$He detector triplets for all five neutron energies chosen for BIFROST (2.7~meV, 3.2~meV, 3.8~meV, 4.4~meV, 5.0~meV) in each Q-channel. 
According to the prismatic analyser concept, each of the three detectors of a triplet records a slightly different region of energy, as neutrons with different energies are scattered in slightly different directions. 

In order to provide enough space for the detector tubes, the analysers and corresponding detectors in adjacent Q-channels are slightly shifted radially. 
The sample--analyser distances are increased or decreased by 4.6--7.5\% in two out of three Q-channels, however the analyser-detector distances are kept unchanged to keep the detectors of same energy on the same vertical planes and by that keeping the spectrometer tank geometry simple.
As a result, the sample--analyser distance is shorter or longer than the analyser--detector distance in two out of three Q-channels, showing a slight asymmetry to the Rowland-geometry.
The three different types of Q-channels are repeated three times, giving the nine Q-channels a ``triple stagger'' geometry.

\begin{figure}[!h]  
  \centering
  \begin{subfigure}{0.6\textwidth} 
      \includegraphics[width=\textwidth]{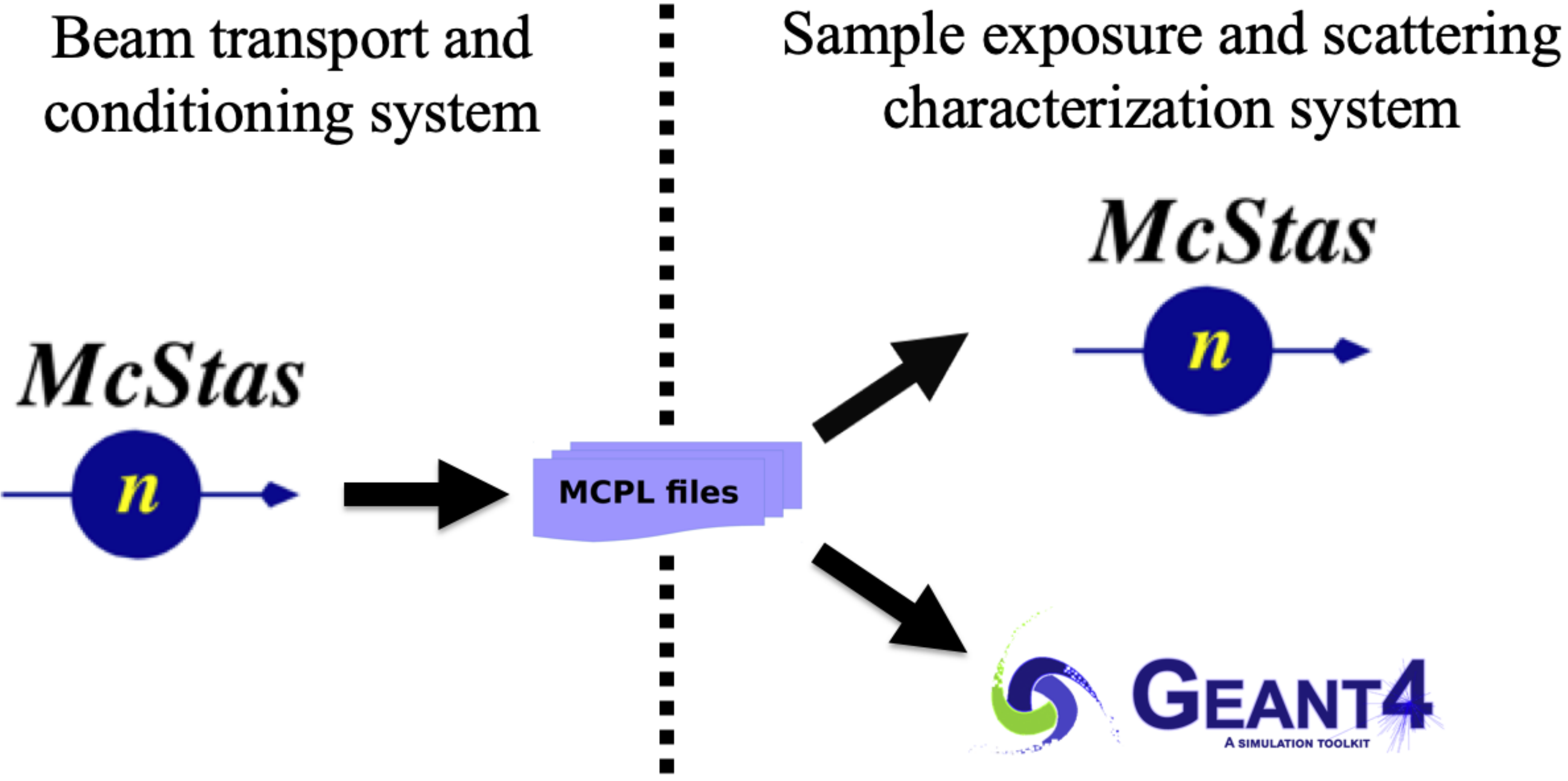}
     \end{subfigure}
   \caption{\footnotesize Outline of simulation scheme.}
  \label{fig:simulationSchema}
\end{figure}

The simulation of the BIFROST instrument is divided into two parts: the simulation of the long beam transport and conditioning system from the moderator until the end of the last guide section before the sample, and the simulation of the sample and scattering characterisation system together (see Fig.~\ref{fig:simulationSchema}). The first part is done using McStas only; the second part is implemented and simulated in both McStas and Geant4, in order to compare the results of these simulation tools and to demonstrate why it is advantageous to use Geant4 for the back end of the instrument. 

The transition between the two parts is facilitated by the aforementioned MCPL tool, that is a binary file format dedicated for storage and interchange of particles between various Monte Carlo simulation applications, like McStas, Geant4, McXtrace~\cite{mcxtrace} and MCNP~\cite{mcnp6}.
For the simulation of neutron transport in crystalline materials, NCrystal is used in both McStas and Geant4.
In the next subsections all simulation tools and models are introduced.

\FloatBarrier
\subsection{McStas model \label{subsec:mcstas}}

McStas~\cite{mcstas1, mcstas2} is a Monte Carlo simulation tool dedicated for simulation of neutron scattering instruments and experiments. 
It is user-friendly, cross-platform, open source and uses a ray-tracing algorithm which enables fast neutron transport simulations over long distances and through many components, that is necessary for long instruments like BIFROST.

\begin{figure}[!h]  
  \centering
  \begin{subfigure}{0.95\textwidth} 
      \includegraphics[width=\textwidth]{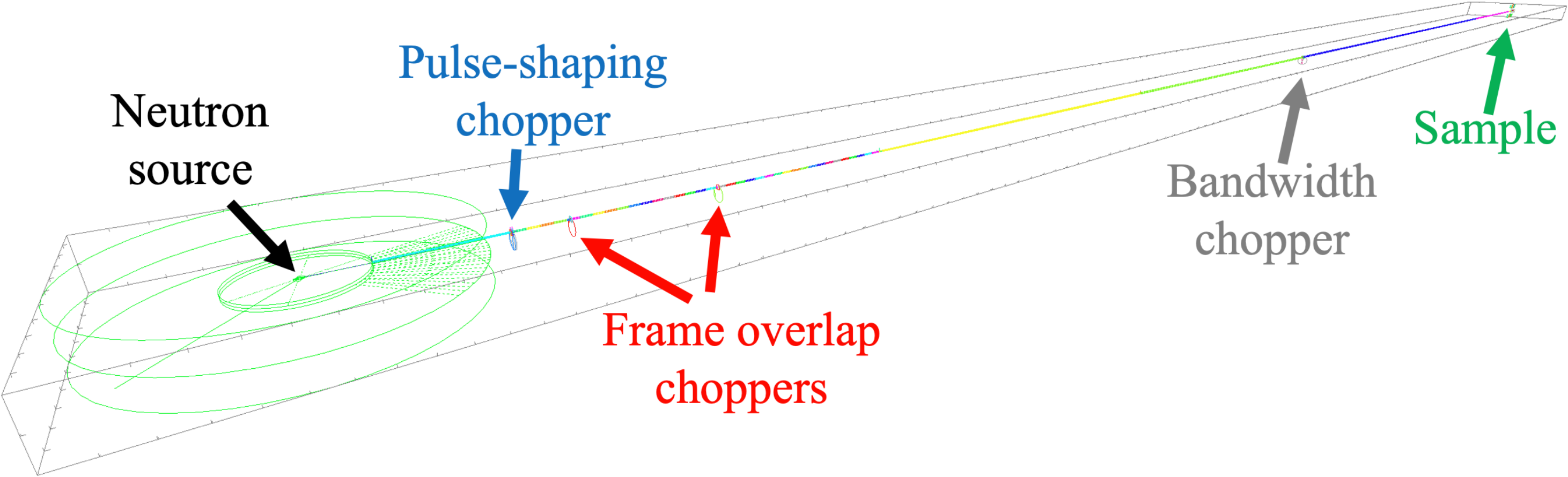}
       \caption{}
     \end{subfigure}
     \begin{subfigure}{0.95\textwidth} 
      \includegraphics[width=\textwidth]{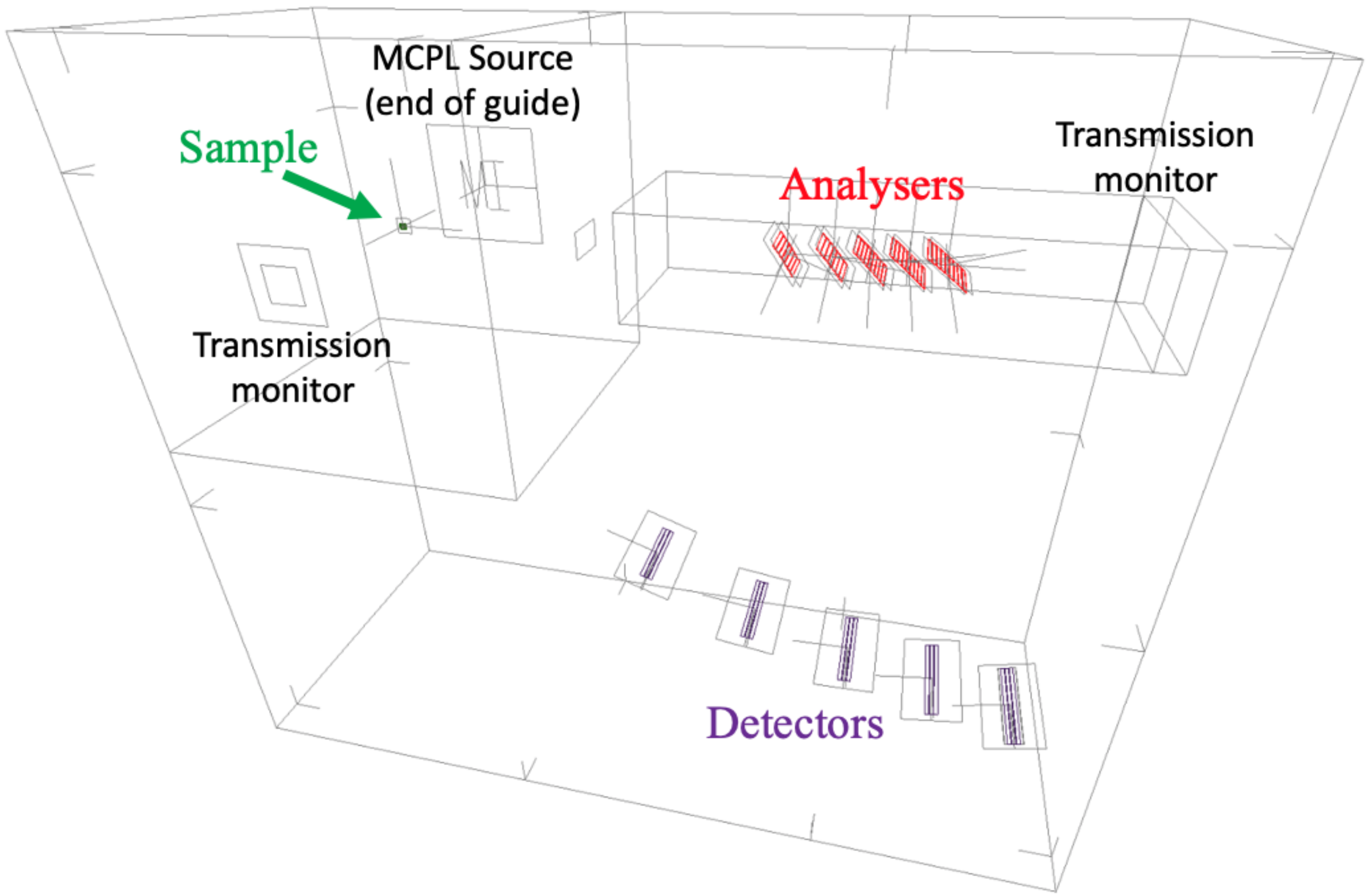}
       \caption{}
     \end{subfigure}
 \caption{\footnotesize McStas model of the beam transport and conditioning system (a) and the sample and scattering characterisation system (b). The figures are at different scale. The neutron source to sample distance is 162~m, the sample to analysers distance is 1189--1622~mm.}
  \label{fig:mcstasModel}    
\end{figure}

For the simulation of neutrons from the source to the end of the beam transport and conditioning system, a previously developed available McStas model of the instrument is used \cite{bifrostMcstasRepo}.
This model, depicted in Fig.~\ref{fig:mcstasModel} contains the butterfly moderator source (``ESS\_butterfly'')~\cite{essmoderator, Zanini_2018}, the four choppers, all guide sections, and several McStas monitor components to characterise the beam at multiple locations along the guide. 
The source is used with the highest intensity, so the deduced rate numbers in this paper correspond to the maximum accelerator power of 5~MW. Expected rates will scale linearly with source power for constant proton energy.
At the end of the last guide section all neutron data are saved in an MCPL file using the ``MCPL\_output'' McStas component. This file serves as input for both the McStas and the Geant4 simulation model of the second part of the instrument.

The McStas model of the sample and scattering characterisation system, depicted in Fig.~\ref{fig:mcstasModel}, contains a crystalline sample, one Q-channel including all five analyser arcs, and McStas monitor components at several places probing ToF, energy and position distribution of neutrons, in order to examine the change of the neutron beam.  
The analyser arcs consist of 7--9 blades using the ``NCrystal\_sample'' component with pyrolytic graphite material, described in more detail in section~\ref{subsec:ncrystal}.  
NCrystal is also used for all crystalline samples throughout this study.
The simulation model does not contain the sample environment, eight out of the nine Q-channels, the filtering system, cross-talk shielding and the detectors. 

Using a reduced geometry and excluding any model of the sample environment is intentional, aiming to get a conservative estimate in terms of the highest detector rates, but implementing only one Q-channel is the result of a limitation coming from the linearity of the McStas simulation process.
In a McStas instrument definition file, the geometrical components like the source, guide sections, choppers, slits and the sample are placed one after the other.
McStas by default propagates neutrons from component to component in the exact order as they appear in this file. 
All neutrons that miss or do not interact with the component downstream are removed from further simulation.
This process makes the simulation of long instruments fast, but on the other hand restricts the neutrons to follow one exact path, that does not allow the simulation of multiple Q-channels simultaneously.
It is possible to change this behaviour by grouping components together, as described in the users and programmers guide~\cite{mcstasmanual}.
This way it is possible to some extent to split the beam, by having a group of components as the potential next target of neutrons, but after the interaction with one of the group members, McStas tries to propagate all neutrons to the same component that appears next in the instrument file.  
This means that if a user wants to split the beam and allow propagation in multiple directions through different components, consecutive groups must be implemented, all of which include the subsequent component in each direction. 
Extensive use of such grouping makes the instrument file immensely complex and still prohibits multiple interactions within one group, or back and forth propagation between groups of components.

For these reasons, only one Q-channel is implemented within which this technique is used to handle the five sets (arcs) of analyser blades, all of which divide the beam into the partition that is scattered towards the detectors and the partition that is propagated towards the next set of analysers (or the beam stop behind the last set).
In order to allow neutrons to proceed without interaction with a set of analyser blades an extra virtual component is added to each group, that mimics an interaction without changing the neutron state, and thereby prevents neutrons to be removed from the simulation. 
As mentioned, neutrons still have to follow the order of the groups, so back-scattering or multiple scattering among blades of the same arc is still not possible in the simulations.

Although the cross-talk shielding between energy- and Q-channels is not explicitly included in the model, as a consequence of the above described process,
a neutron can reach a particular detector tube only by scattering in one of the corresponding analyser blades. This is practically equivalent to an ideal cross-talk shielding absorbing all stray neutrons. 
The case is similar for the filtering system, that is replaced by a monitor component, that transmits all neutrons below 7~meV energy and none above.
As mentioned earlier, the transmission of beryllium drops sharply around 5~meV, that is in fact the highest of the five final energies selected by the analysers.
Simulation of effects of this transition in the transmission is out of the scope, and using ideal transmission in the 0--7~meV energy region keeps the rate estimates conservative.

The intent is to determine the incident detector rates, therefore the simulation of the detection process is out of scope. 
Detectors are modelled with McStas monitor components, and a neutron is counted as incident for a detector tube if it crosses the plane at the centre of the detectors within the outline of that particular tube. The sample--analyser distance is equal to the analyser--detector distance, meaning that symmetrical Q-channels are modelled. 
 
As this subsection demonstrates, using McStas to model such complex system as the analyser-detector system of BIFROST in detail is not straightforward and is subject to certain limitations.

McStas version 2.5 is used for the simulations.

 \FloatBarrier
\subsection{Geant4 model}
 
Geant4~\cite{geant4a,geant4b,geant4c_inpresscorrectedproof} is a general purpose Monte Carlo particle transport toolkit developed at CERN with applications in many fields, e.g.\,high energy physics, nuclear physics, accelerators and medical physics.
Its usability for simulation of neutron detectors has been greatly improved by the ESS Detector Group by building a framework~\cite{KANAKI2018386, dgcodechep2013} around it, which adds several functionalities and integrates NCrystal and MCPL. 

\begin{figure}[!h]  
  \centering
  \begin{subfigure}{0.9\textwidth} 
       \includegraphics[width=\textwidth]{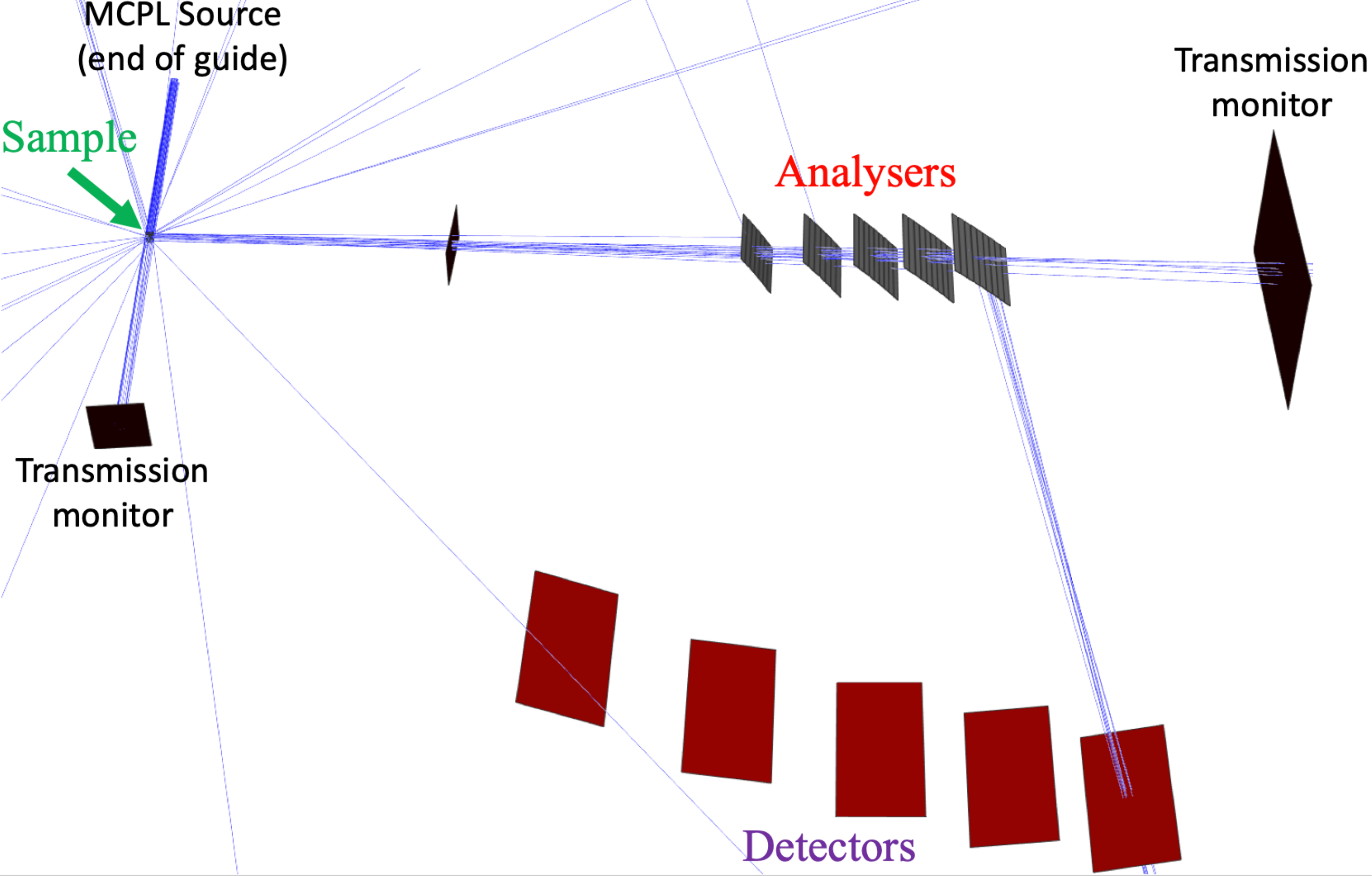}
        \caption{}
     \end{subfigure}
     \hfill%
     \begin{subfigure}{0.9\textwidth} 
       \includegraphics[width=\textwidth]{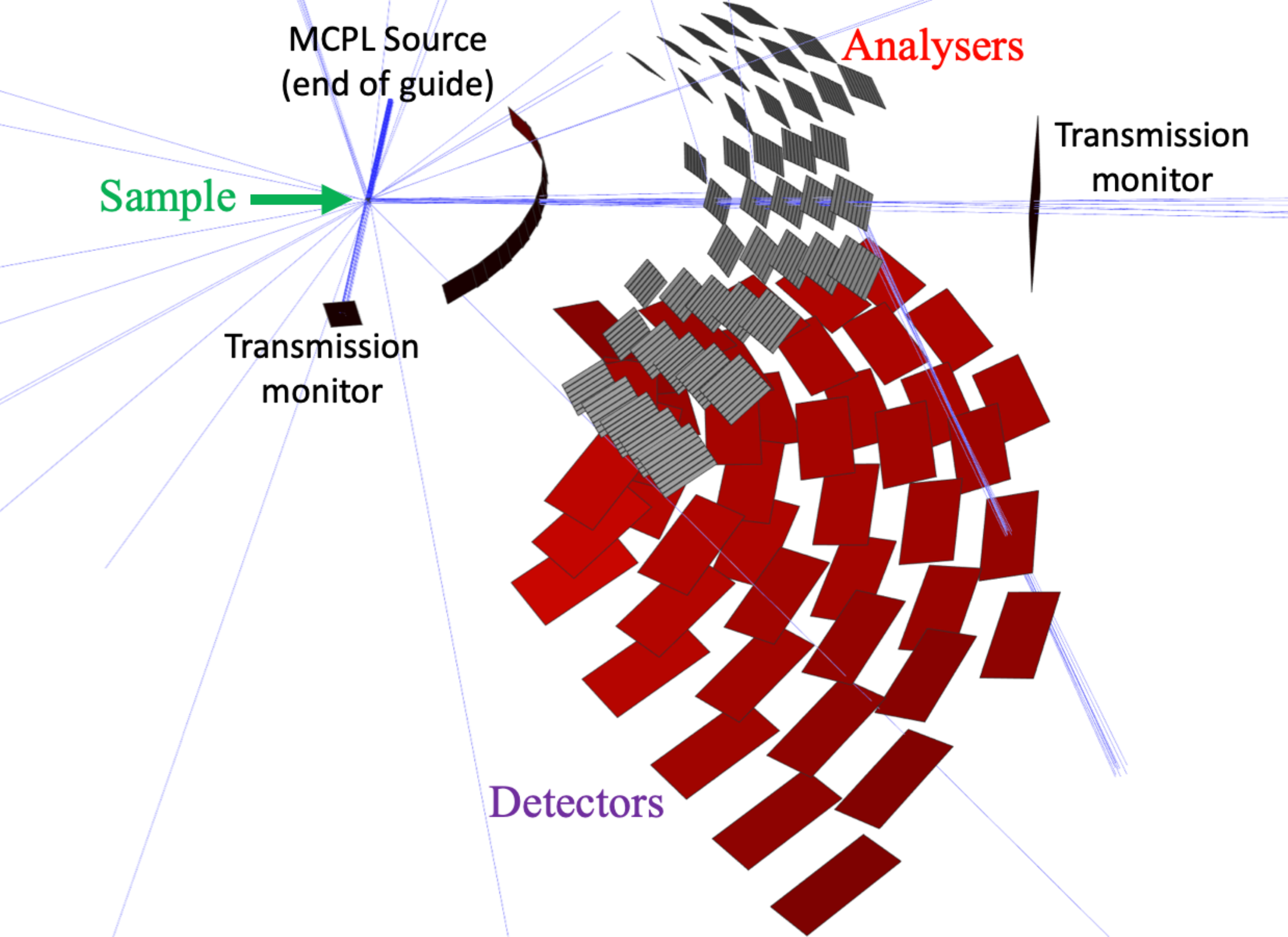}
 \caption{}	
     \end{subfigure}
   \caption{\footnotesize Geant4 model of the sample and scattering characterisation system including one Q-channel (a) or all Q-channels (b). The blue lines indicate simulated neutron paths.}
  \label{fig:geantModel}
\end{figure}

The Geant4 simulation model of the sample and scattering characterisation system, depicted in Fig.~\ref{fig:geantModel}, contains the same parts as the McStas model (crystalline sample, all five analyser arcs in a Q-channel) but with the option to simulate with all Q-channels included. 
To make the results comparable with the McStas model, the McStas monitor components are mimicked with empty volumes with exactly the same location and surface, in order to create histograms with the same predefined spatial, energy and time-of-flight (ToF) resolution.

Although the cross-talk shielding could be easily implemented in the model, for the same comparison purposes it is replaced by certain conditions applied at data in the analysis level. This means that neutrons cannot skip parts of the model. 
They can, however, scatter back and forth between the geometrical components many times, unlike in McStas.
This gives the possibility to analyse effects of cross-talk on signals and to evaluate shielding strategies.

The Geant4 physics list used is QGSP\_BIC\_HP\_EMZ that uses high precision models and cross-sections for neutron energies lower than 20~MeV, and allows the correct treatment of thermal and cold neutrons when combined with NCrystal.
 
 \FloatBarrier
\subsection{NCrystal \label{subsec:ncrystal}} 

NCrystal~\cite{KANAKI2018386, CAI2020106851, ncrystalManuscript} is a novel open source software package for modelling thermal neutron transport in crystalline materials.
It consists of a data library and associated tools which enables calculations for Monte Carlo simulations. 
It can be used together with McStas and Geant4, to enhance their capabilities of the correct treatment of neutron transport in typical components of neutron instruments, including beam filters, monochromators, analysers, samples and detectors.
Physics modelled by NCrystal includes both coherent elastic (Bragg diffraction) and incoherent or inelastic (phonon) scattering.
It treats all valid Bragg diffractions on each reflection plane explicitly and is able to use various models for inelastic scattering on phonons.
Its data library~\cite{ncrystalDataLibrary} already contains the most popular crystals and the results are validated against the EXFOR database and existing crystallographic software.

NCrystal focuses initially on scattering in single-crystals or polycrystalline materials and powders.
Most single crystalline materials are appropriately modelled with crystallites orientated around some reference orientation with a Gaussian distribution that has a standard deviation of the mosaicity of the crystal.
There are, however, single crystalline materials with crystallite distributions so different from Gaussian that this approximation does not hold.
One of these materials is pyrolytic graphite, that is widely used as a monochromator and analyser in neutron instruments.
This is precisely the case for BIFROST, where 369 highly oriented pyrolytic graphite analyser blades are used altogether in the nine Q-channels.
Graphite has a layered structure, made up of graphene sheets in which carbon atoms are arranged in a hexagonal lattice.
In highly oriented pyrolytic graphite the crystallite axes orthogonal to the graphene sheets are distributed along a preferred direction, suitable
for description with a Gaussian mosaicity distribution, but the orientation around this axis is completely random, resulting in powder-like features in neutron scattering.
Recent developments enabled NCrystal to handle materials with this kind of structure by using a specialised model for layered crystals.

\begin{figure}[!h]  
  \centering
  \begin{subfigure}{0.49\textwidth} 
      \includegraphics[width=\textwidth]{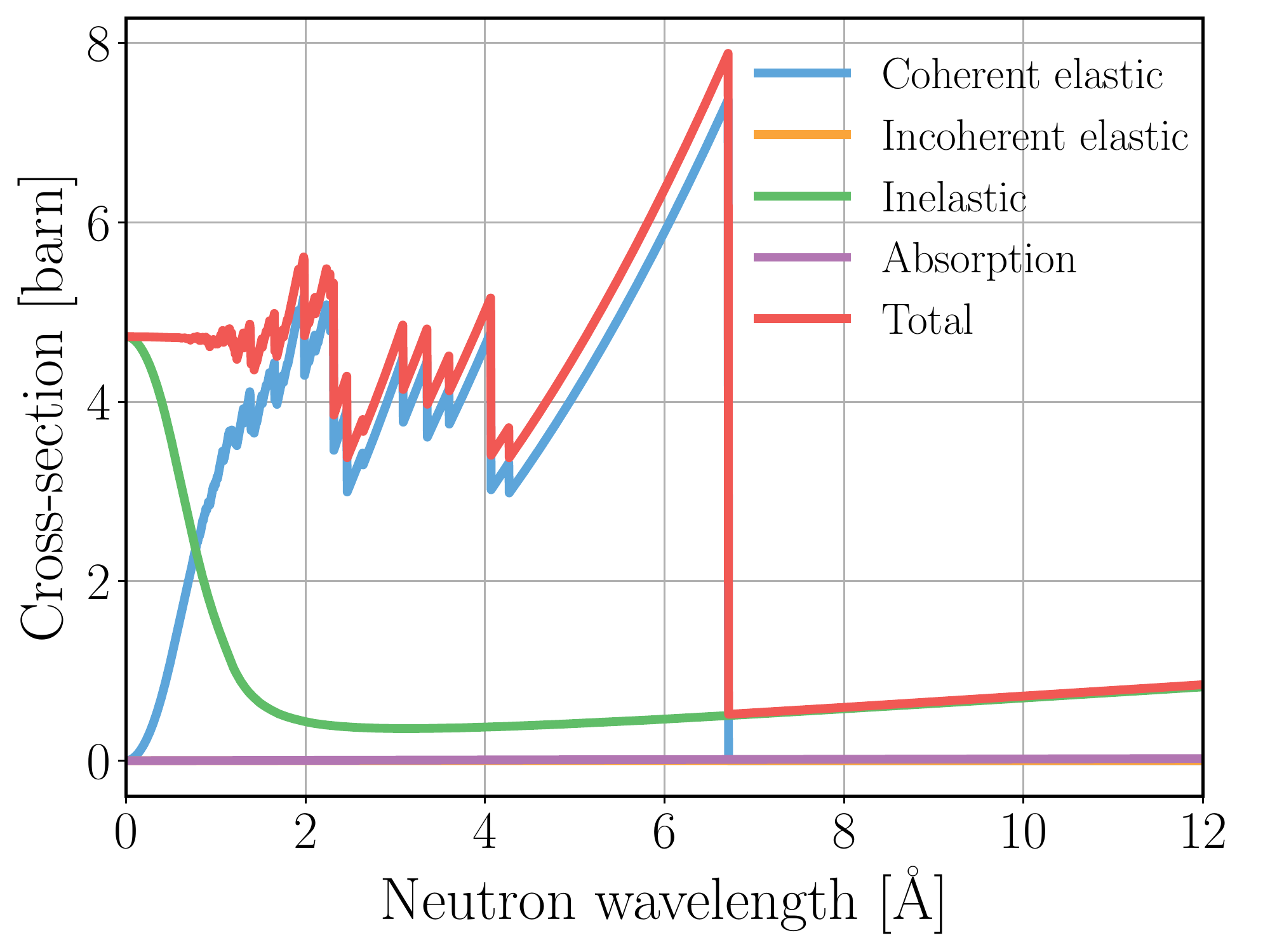}
     \end{subfigure}
     \begin{subfigure}{0.49\textwidth} 
      \includegraphics[width=\textwidth]{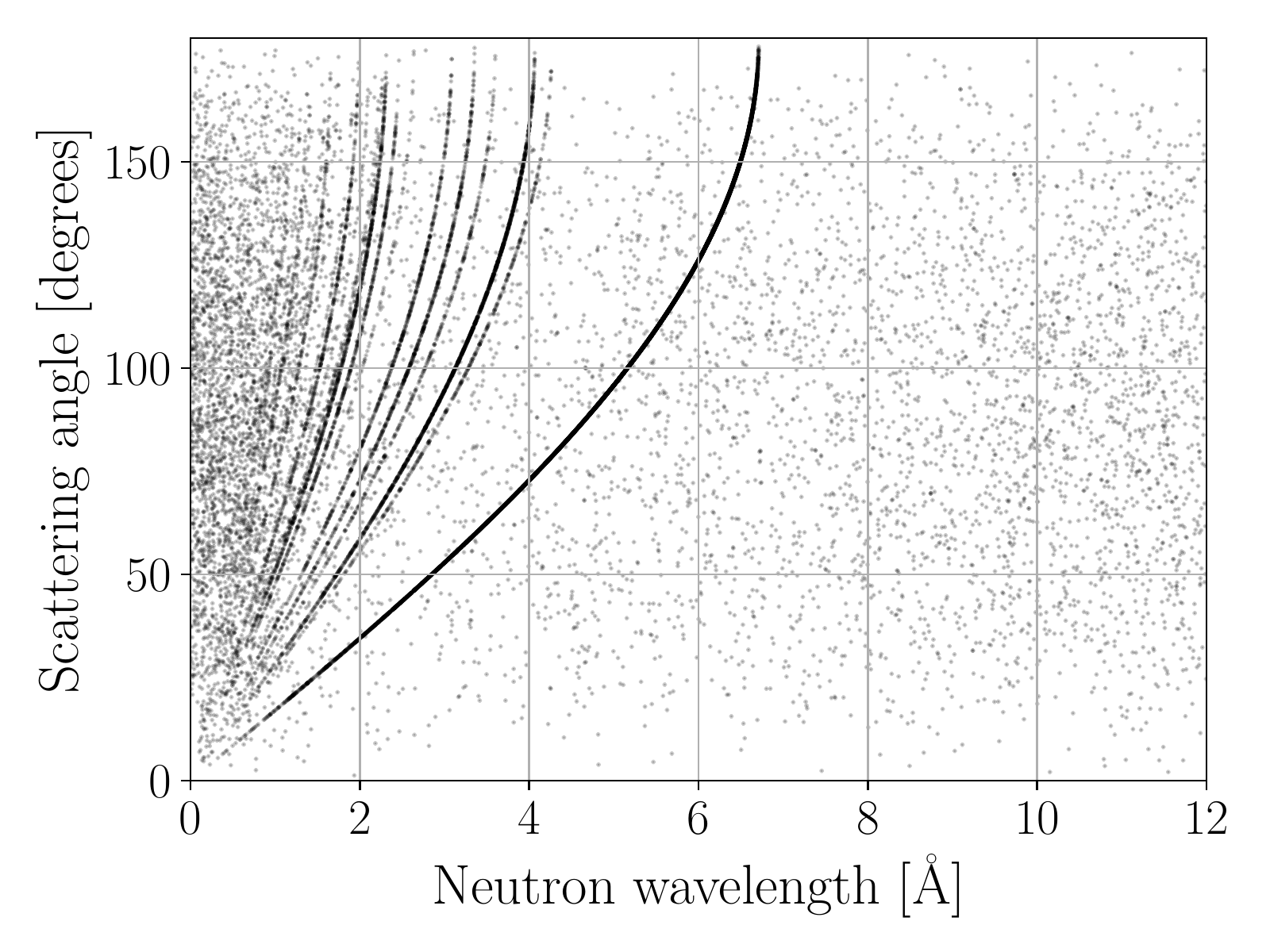} 
     \end{subfigure}
  \caption{\footnotesize Components of the total interaction cross-section (a) and two-dimensional scatter-plot (b) for NCrystal pyrolytic graphite powder.}
  \label{fig:NCrystalCrossSection}   
\end{figure}

NCrystal is used for the crystalline materials both in the McStas and Geant4 model of BIFROST. 
The samples are modelled as single-crystals, but for the pyrolytic graphite analysers the new layered crystal model is used.
Demonstrating some of the NCrystal pyrolytic graphite properties, the components of total cross-sections and a distribution of randomly sampled scatter angles are depicted in Fig.~\ref{fig:NCrystalCrossSection} for powder sample, not the layered crystal distribution. As presented in the NCrystal data library~\cite{ncrystalDataLibrary}, the cross-sections are validated against experimental data~\cite{NSR1960WA07}.

\begin{figure}[!h]  
  \centering
  \begin{subfigure}{0.7\textwidth} 
      \includegraphics[width=\textwidth]{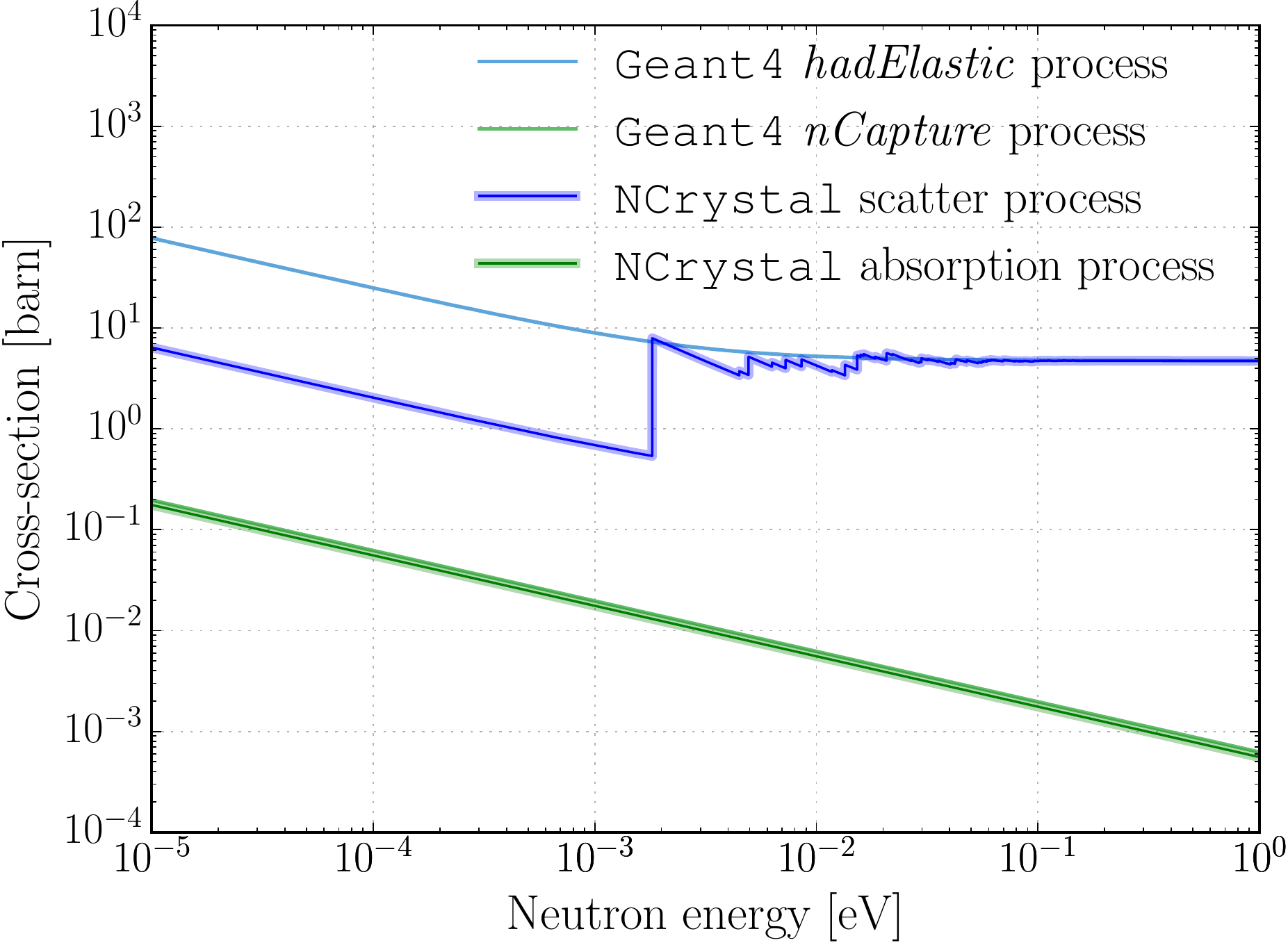}
     \end{subfigure}
   \caption{\footnotesize Comparison of neutron interaction cross-sections of pyrolytic graphite powder in Geant4 and NCrystal. Lines for Geant4 neutron capture and NCrystal absorption cross-sections are barely distinguishable, the former being higher in the whole energy range depicted.}
  \label{fig:compareCrossSection}
\end{figure}

Up until the latest version (v1.0.0), NCrystal treats absorption with the simple model of absorption cross-sections being inversely proportional to the neutron velocity.
The absorption cross-section for a particular neutron velocity is calculated by scaling the value given at the reference velocity of 2200~m/s. 
This applies for NCrystal materials used in McStas simulations, but not for those in Geant4 simulations. 
Geant4 models secondary particles produced in absorption, therefore the NCrystal plugin does not interfere with the Geant4 absorption physics at all.
As an example, the minor differences in absorption cross-section for pyrolytic graphite is shown in Fig.~\ref{fig:compareCrossSection}.

All results of simulations have systematic and statistical uncertainties.
Unlike systematic uncertainties which are present mainly due to imperfect modeling of the system,
statistical uncertainties can be reduced by increasing the number of simulated particles. 
Throughout this work where the uncertainty is not significant, it is not indicated.
Sources of systematic uncertainties are considered and not expected to change the conclusions.

With the simulation models and tools described in this section, a detailed analysis is carried out, in order to give an estimate of the incident rates that are anticipated for detectors at BIFROST. 
The incident detector rates for elastic peaks using various samples and instrument parameters are presented in the next section, as well as a comparison of results with McStas and Geant4 simulations.

\FloatBarrier \clearpage
\section{Incident rates for coherent elastic peaks}

Determination of anticipated detector rates for an instrument is a key part of defining requirements for the detectors to be used.
It can prevent the detector rate capability from becoming the bottleneck of experiments or a source of performance degradation.
For this reason, the simulation tools and models described in the previous section are used to determine the highest time-averaged and the highest instantaneous (peak) incident rates for the detector tubes. These rates are determined for the highest-case and for more realistic operational conditions. 
 
The highest-case incident rate for a single detector tube occurs when a strong Bragg peak from a single crystal sample gets reflected to it.
To get neutrons scattered on the sample onto the detectors, their energy has to match one of the energies selected by the analyser arcs.
Based on the ESS source spectum, the guide transmission and the energy resolution of the analysers, the highest incident rates are expected for the 5.0~meV (4.045~\AA) neutrons. 
 

Regarding highly reflective materials that would result in the highest detector rates,
the truly highest-case sample would be a pyrolytic graphite single crystal (d$_{002}$=3.3555~\AA), but to get results from a less unrealistic sample with large enough lattice parameter and strong Bragg peak, simulations are also done using an yttrium oxide (Y$_2$O$_3$) single crystal (d$_{2 -2 -2}$=3.0724~\AA).
 
Further parameters that influence the rates on detectors are the pulse-shaping chopper opening time, sample size, and sample mosaicity. 
 To get the highest possible rates, maximum flux mode is used, when the pulse-shaping chopper is fully open, resulting in a 10$^{10}$~n/s/cm$^2$ flux on sample.
The instrument is designed to facilitate measurements on small samples, but dimensions up to 1.5~cm are possible, therefore cylindrical samples with the diameter and height of 1.5~cm are used in the simulations.
The sample mosaicity resulting in the highest rate can depend on the divergence of the incident beam on the sample, however, as a rule of thumb the highest rates are expected when the mosaicity of the sample matches that of the analysers, so the sample mosaicity is set to 60~arcmin. 
It is shown later in subsection~\ref{subsec:sample_mosaicity} that this is a good assumption, and within the $60\pm$20~arcmin sample mosaicity range it has a less than a 10\% effect on the incident detector rates. 
  
In order to realise the simulations, the samples are oriented to fulfil the Bragg condition for the incoming 5~meV neutrons on the selected scattering planes, and the single Q-channel modelled is rotated according to the resulting scattering angle. 
In BIFROST the whole scattering characterisation system can be rotated around the sample, so having a Bragg peak from any sample in the exact direction of a single Q-channel is perfectly realistic.

The simulation with the parameters described above is done in two steps. 
First, the neutron transport from the source to the end of the beam transport and conditioning system is done with the McStas model, saving neutron data at the end in an MCPL file.
This file is then used as the source term for the simulation of the sample and scattering characterisation system, that is done with both McStas and Geant4. The results are presented in the following subsection (\ref{mcstasGeantComparison}). In the subsequent subsection the impact of different parameters like sample and analyser mosaicity, sample size, pulse-shaping chopper opening time is scanned, but due to the good agreement of McStas and Geant4 simulations results (demonstrated later in subsection~\ref{mcstasGeantComparison}), this is only done using the Geant4 model.

 \FloatBarrier
\subsection{McStas -- Geant4 comparison \label{mcstasGeantComparison}}

\subsubsection{Pyrolytic graphite sample}

\begin{figure}[!h]  
  \centering
  \begin{subfigure}{1.0\textwidth} 
      \includegraphics[width=\textwidth]{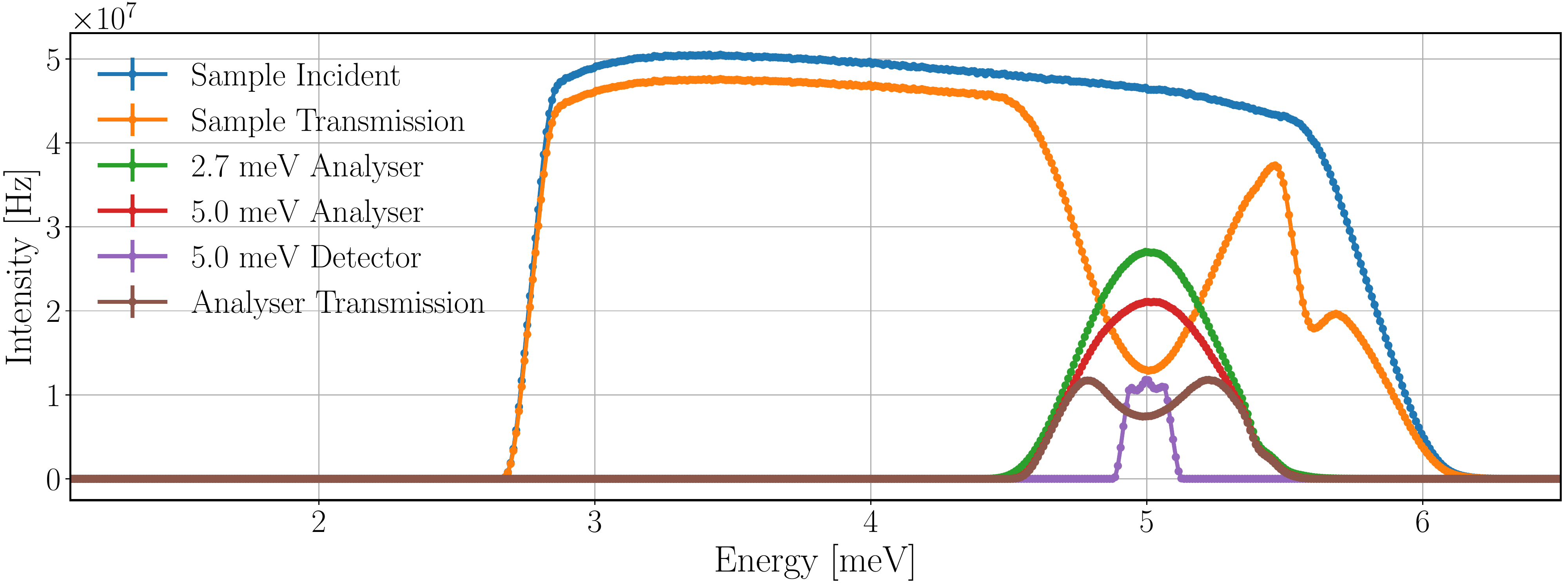}
     \end{subfigure}
      \begin{subfigure}{1.0\textwidth} 
      \includegraphics[width=\textwidth]{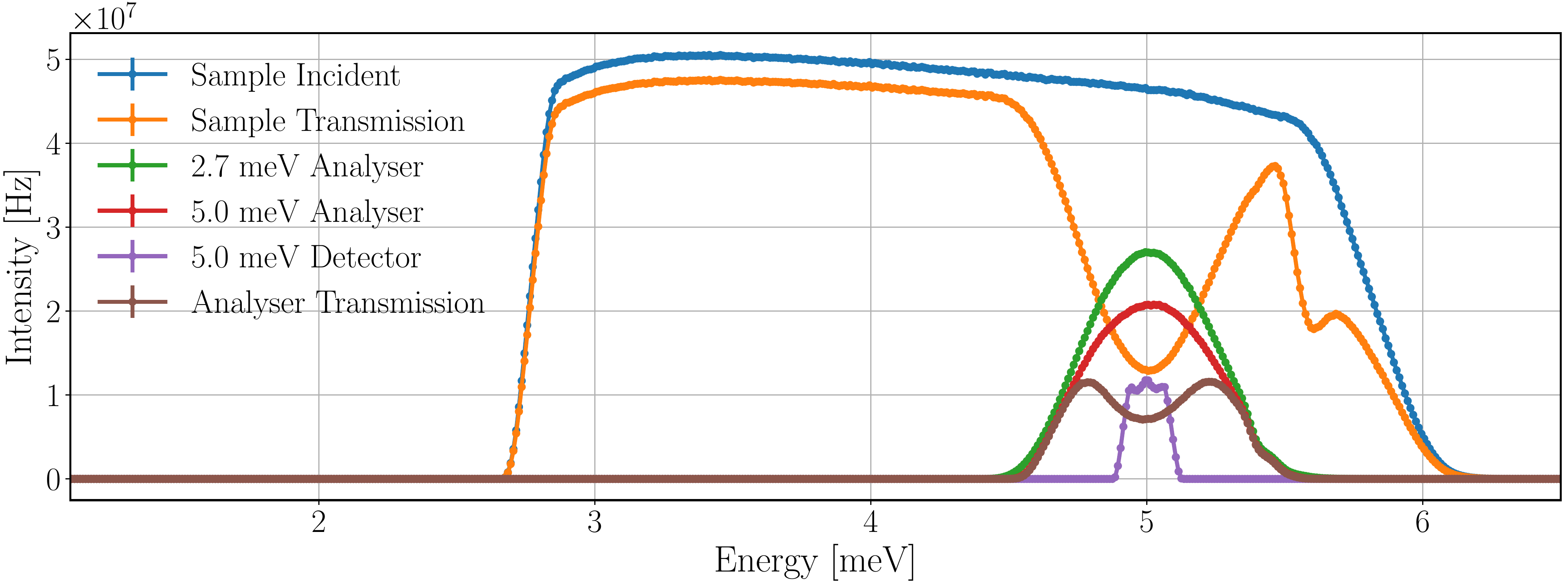}
     \end{subfigure}
   \caption{\footnotesize Time-averaged neutron energy spectra at the sample and the scattering characterisation system with pyrolytic graphite single crystal sample in McStas (a) and Geant4 (b) simulation. Incident beam on sample (in blue), beam transmitted through the sample (in orange), beam on the set of analysers for 2.7~meV neutrons (in green), beam on the set of analysers for 5.0~meV neutrons (in red), neutrons hitting the detector triplet for 5.0~meV (in purple), beam transmitted through all sets of analysers (in brown). The lines are only joining the points.}
  \label{fig:changeOfEnergyPG}
\end{figure}

Fig.~\ref{fig:changeOfEnergyPG} depicts the time-averaged energy spectra of neutrons at the sample and different parts of the scattering characterisation system from the McStas and Geant4 simulations using a pyrolytic graphite single crystal sample.

Neutrons of a broad energy range -- centred around 5.0~meV -- are scattered on the sample toward the Q-channel, reaching the 2.7~meV analysers and therefore missing from the sample transmission spectrum.
The change of the spectrum between the 2.7~meV and 5.0~meV analysers is caused by the spread of neutrons, and absorption in the analysers. 
The neutrons selected by the 5.0~meV analysers are scattered towards the corresponding detector triplet, and are therefore absent from the analyser transmission spectrum.

The time-averaged neutron intensities acquired by the integration of the energy spectra are shown in Tab.~\ref{tab:changeOfEnergyPG}.
The results of the McStas and Geant4 simulations agree with only minor differences.

\begin{table}[htp]
\centering
\begin{tabular}{|ccc|}
\hline
Position & McStas [Hz] & Geant4 [Hz] \\\hline
Sample Incident & 1.45$\cdot$10$^{10}$ & 1.45$\cdot$10$^{10}$ \\
Sample Transmission & 1.16$\cdot$10$^{10}$ & 1.16$\cdot$10$^{10}$ \\
2.7~meV Analyser& 1.50$\cdot$10$^{9}$ & 1.50$\cdot$10$^{9}$ \\
5.0~meV Analyser& 1.20$\cdot$10$^{9}$ & 1.19$\cdot$10$^{9}$ \\
5.0~meV Detector & 1.98$\cdot$10$^{8}$ & 1.98$\cdot$10$^{8}$ \\
Analyser Transmission& 7.47$\cdot$10$^{8}$ & 7.31$\cdot$10$^{8}$ \\\hline
\end{tabular}
\vspace{-.1cm}
\caption{\footnotesize Time-averaged neutron intensities at the sample and different parts of the scattering characterisation system with pyrolytic graphite single crystal sample with McStas and Geant4.}
\label{tab:changeOfEnergyPG}
\end{table}

\begin{figure}[!h]  
  \centering
  \begin{subfigure}{0.49\textwidth} 
      \includegraphics[width=\textwidth]{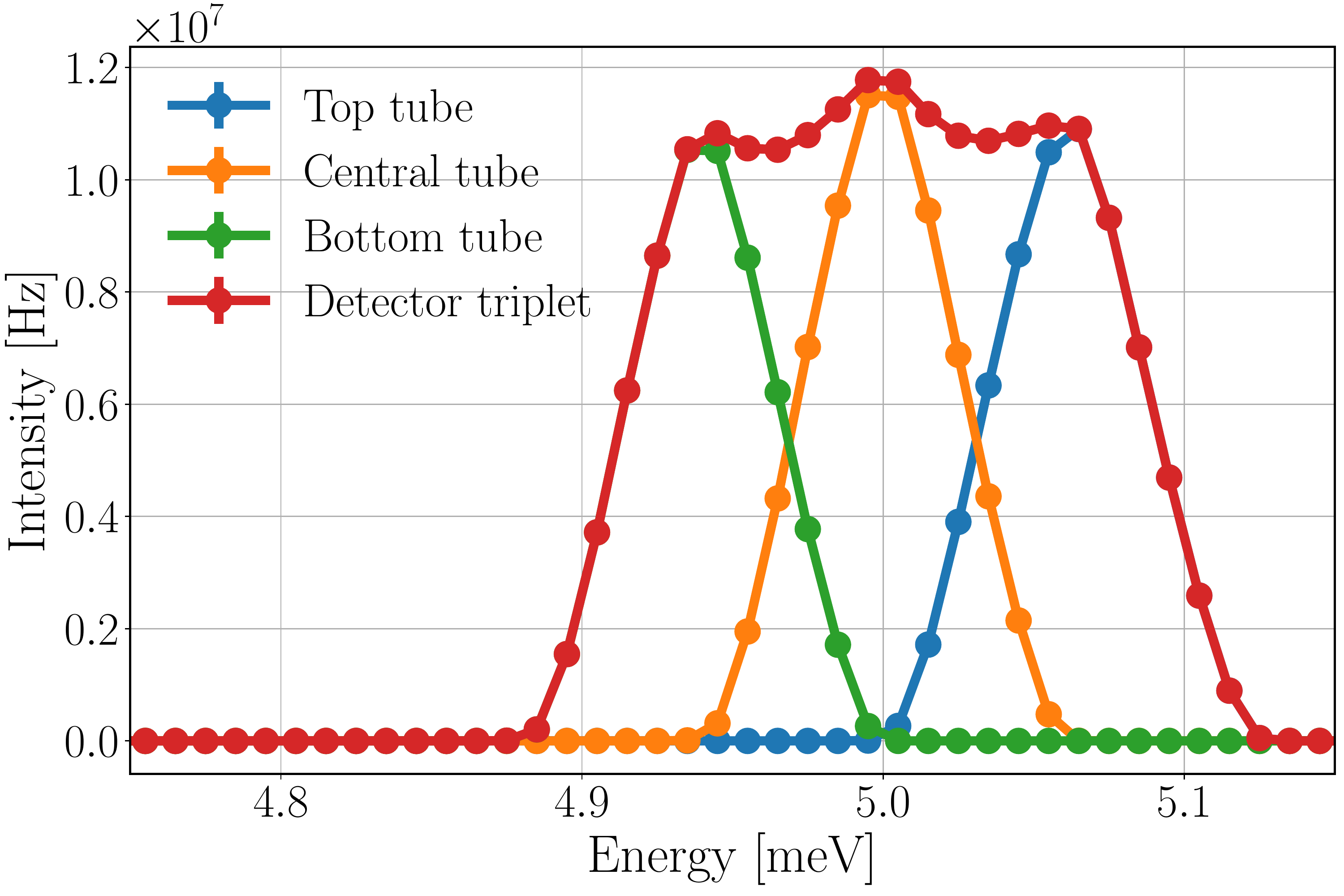}
  \end{subfigure}
   \begin{subfigure}{0.49\textwidth} 
      \includegraphics[width=\textwidth]{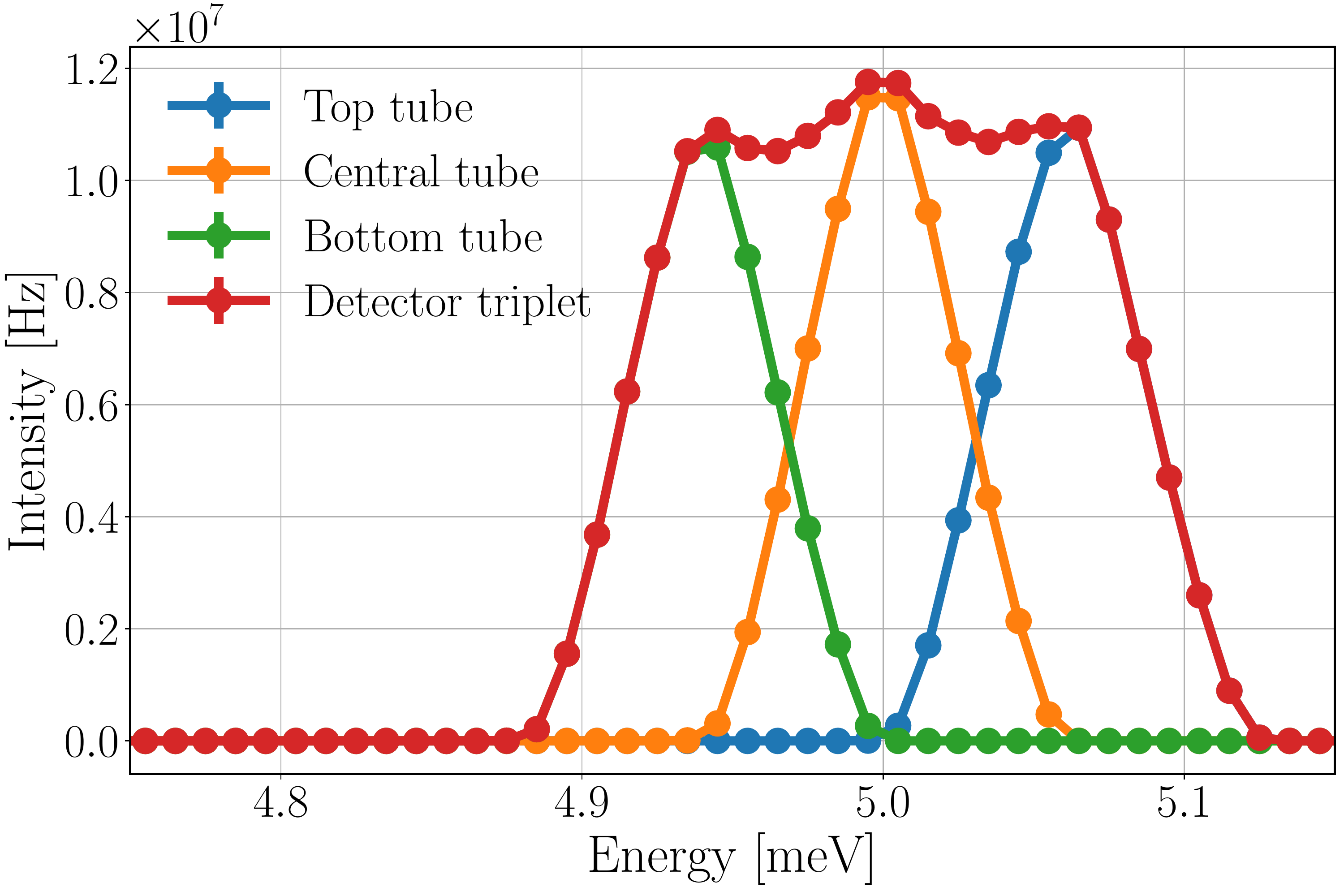}
  \end{subfigure}
  \caption{\footnotesize Time-averaged incident neutron energy spectra of the 5~meV detector tubes in McStas (a) and Geant4 (b) simulations, with pyrolytic graphite sample. The lines are only joining the points.}
  \label{fig:detTripletSpectra}
\end{figure}

\begin{figure}[!h]  
  \centering
  \begin{subfigure}{1.0\textwidth} 
      \includegraphics[width=\textwidth]{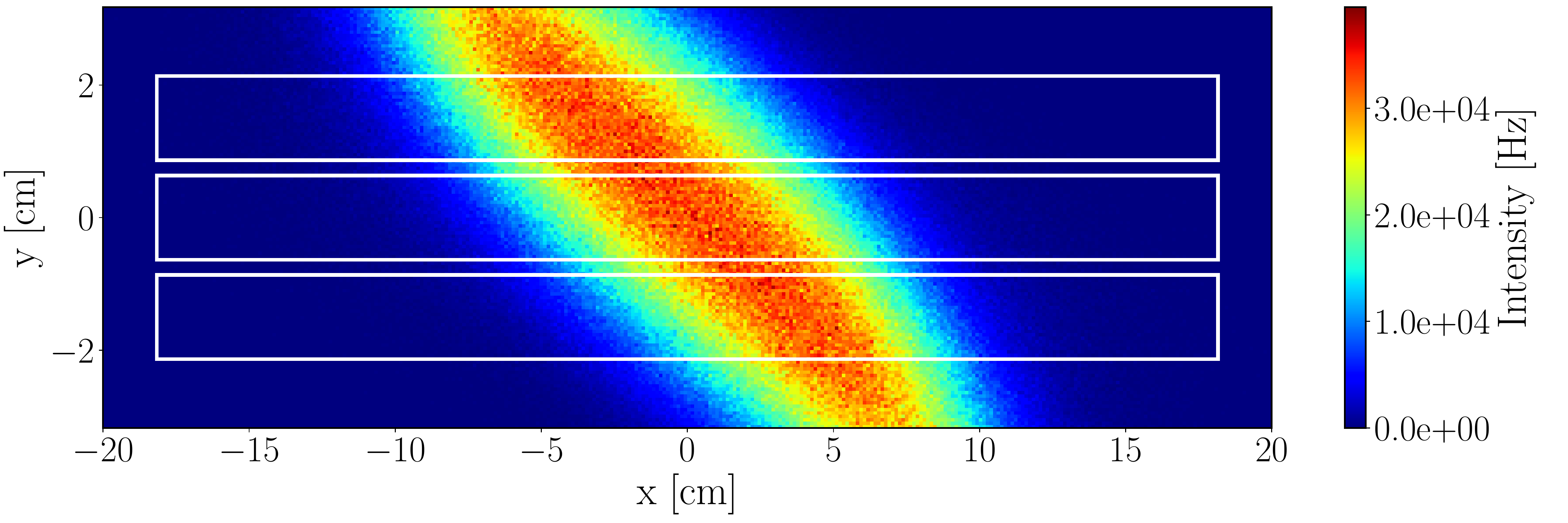}
  \end{subfigure}
  \begin{subfigure}{1.0\textwidth} 
      \includegraphics[width=\textwidth]{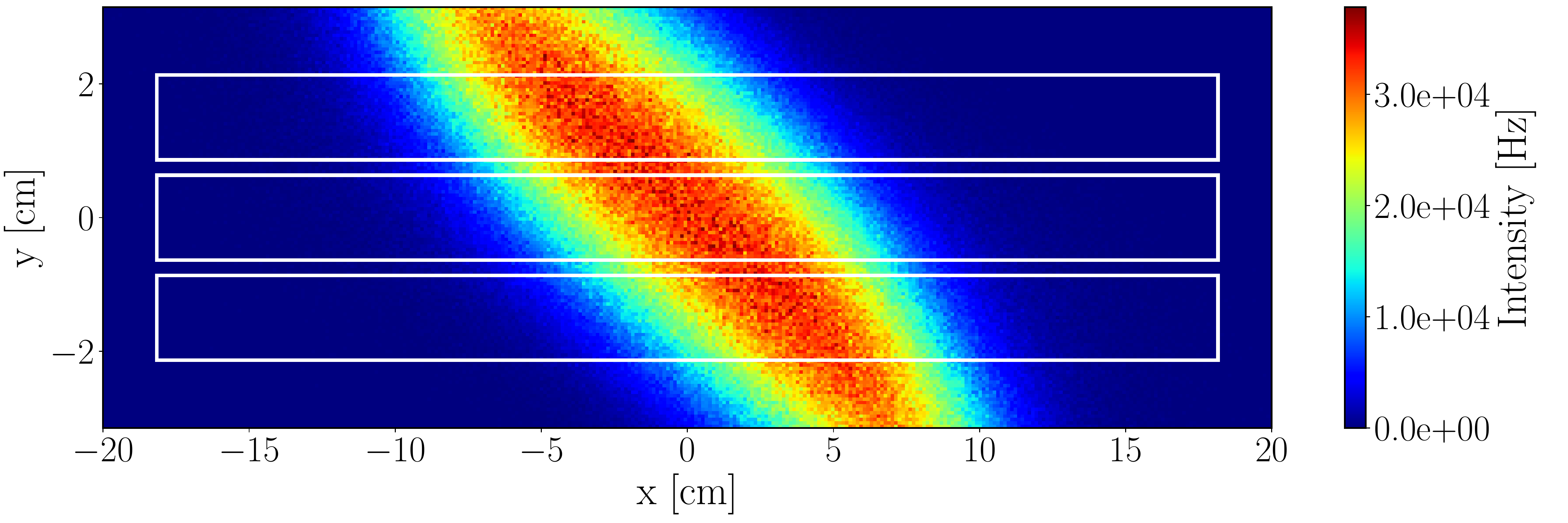}
  \end{subfigure}
  \caption{\footnotesize Time-averaged neutron intensities in the plane of the 5~meV detector tubes in McStas (a) and Geant4 (b) simulations, with pyrolytic graphite sample. The white lines indicate the outlines of the detector tubes.}
  \label{fig:PGtimeAveragePSD}
\end{figure}

The structure of the energy spectrum of the detector triplet is the result of summing the spectra of all three detectors, as depicted in Fig.~\ref{fig:detTripletSpectra}.
As expected, the analysers scatter neutrons with slightly different energies in slightly different directions -- in accordance with the Bragg's law -- and as result of this vertical spread, the three detectors of the triplet record slightly different regions of energy, in accordance with the prismatic analyser concept.
The sample spreads the neutron beam similarly but in the horizontal plane.
The combined effect of these processes is visible in Fig.~\ref{fig:PGtimeAveragePSD}, showing a diagonal shape in the time-averaged neutron intensities in the plane over the detector tubes.

\FloatBarrier

Integrating the incident neutron intensities in Fig.~\ref{fig:PGtimeAveragePSD} over the areas of the tubes gives the time-averaged incident rates for the tubes, presented in Tab.~\ref{tab:PGaverageTubeRates}. The results show that the time-averaged incident rate for a single detector tube can be almost as high as 70~MHz.


\begin{table}[htp]
\centering
\begin{tabular}{|ccc|}
\hline
Detector tube & McStas [Hz] & Geant4 [Hz] \\\hline
Top & 6.69$\cdot$10$^{7}$ & 6.70$\cdot$10$^{7}$ \\
Central & 6.95$\cdot$10$^{7}$ & 6.93$\cdot$10$^{7}$ \\
Bottom & 6.20$\cdot$10$^{7}$ & 6.20$\cdot$10$^{7}$  \\\hline
\end{tabular}
\vspace{-.1cm}
\caption{\footnotesize Time-averaged incident neutron rates of the 5~meV detector tubes in McStas and Geant4 simulations, with pyrolytic graphite sample.}
\label{tab:PGaverageTubeRates}
\end{table}

\begin{figure}[!h]  
  \centering
  \begin{subfigure}{0.49\textwidth} 
      \includegraphics[width=\textwidth]{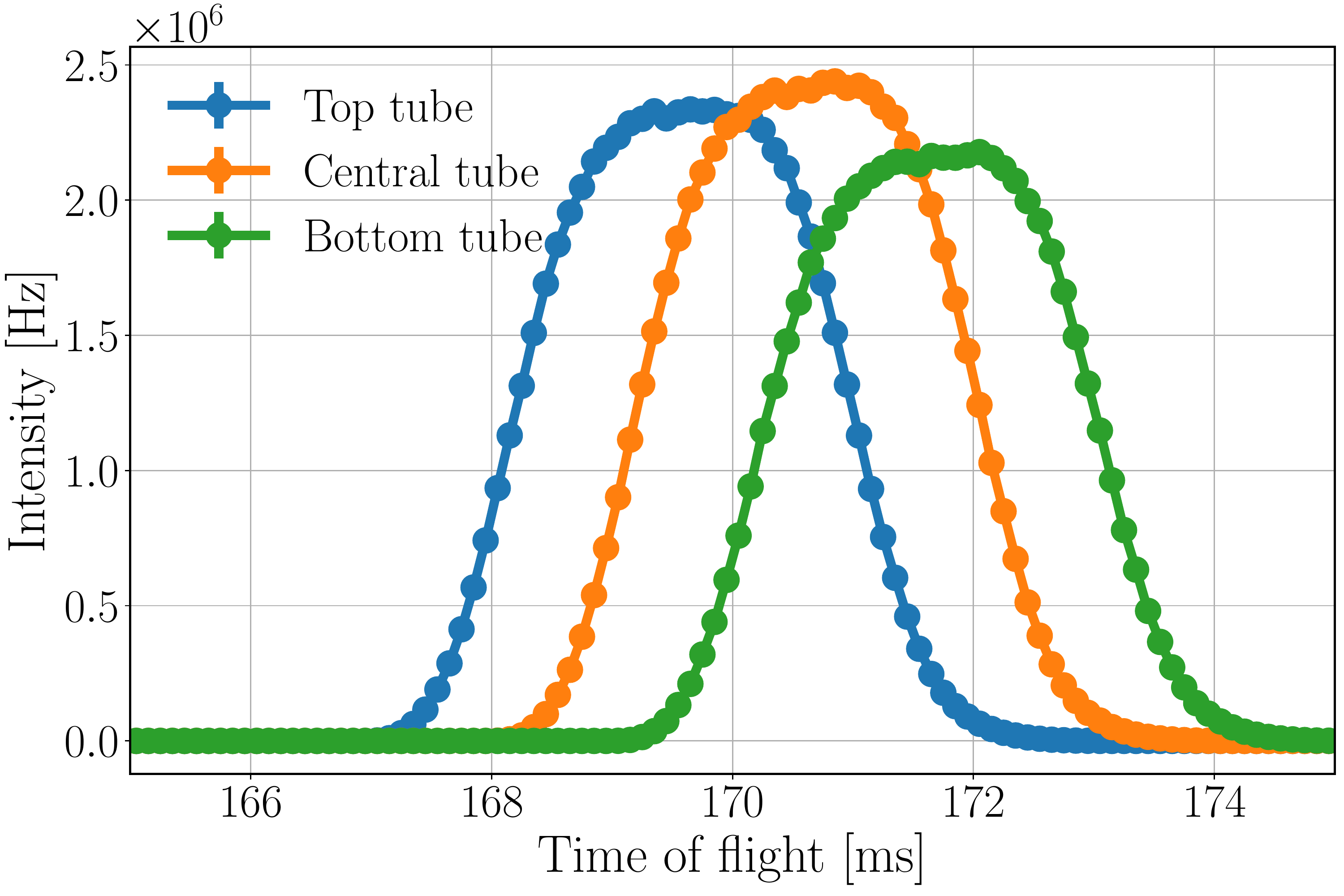}
     \end{subfigure}
     \begin{subfigure}{0.49\textwidth} 
      \includegraphics[width=\textwidth]{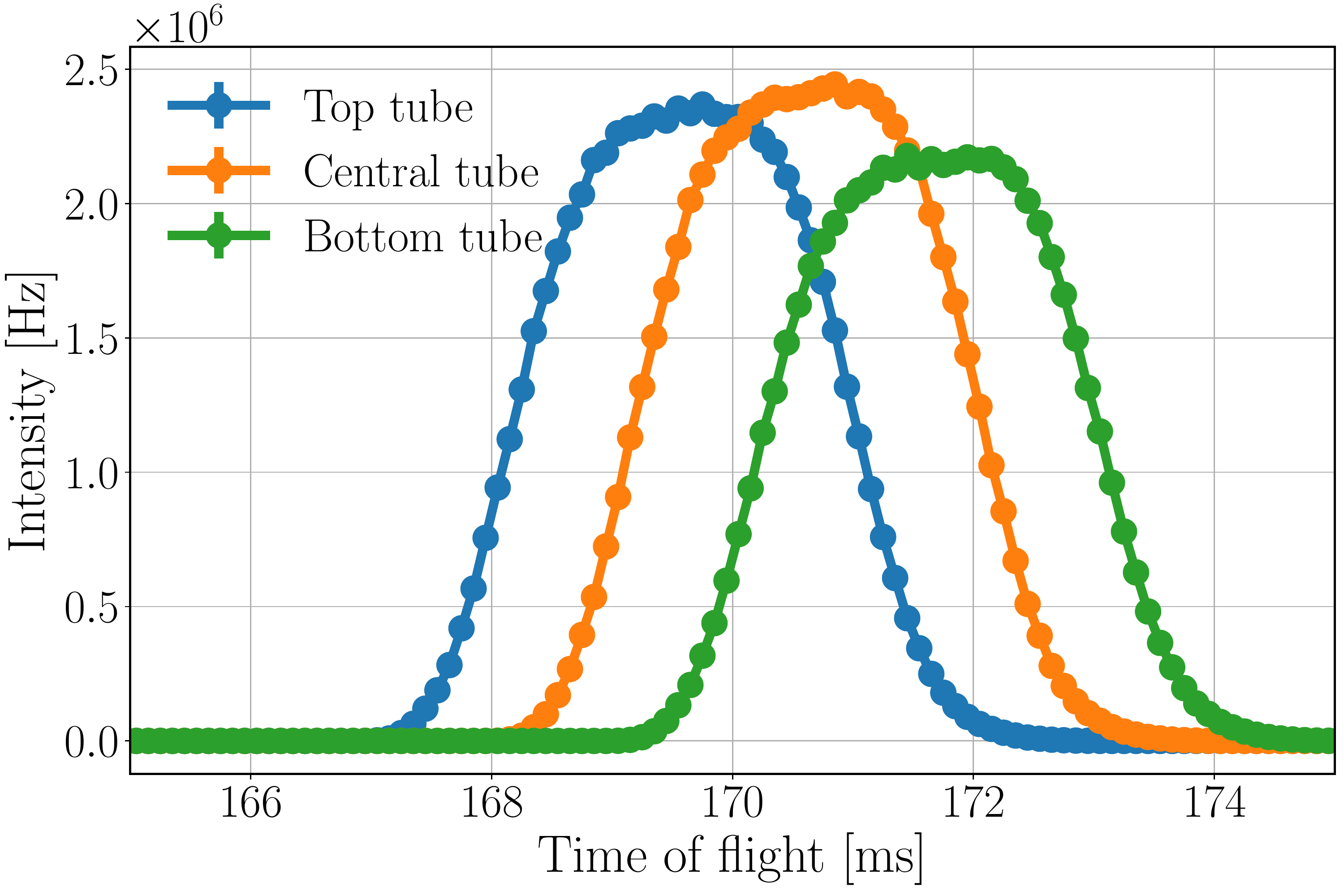}
     \end{subfigure}
   \caption{\footnotesize ToF spectrum of neutrons at the 5~meV detector triplet in McStas (a) and Geant4 (b) simulations, with pyrolytic graphite sample. The lines are only joining the points.}
  \label{fig:tof_spectrum}
\end{figure}

Given that BIFROST is a ToF instrument at a spallation neutron source, the incident detector rate has a pulsed time structure. 
The ToF distribution of a single pulse on the 5~meV detectors is depicted in Fig.~\ref{fig:tof_spectrum}.
By taking into account only those neutrons which arrive to the detectors at the peak of their ToF distribution in a short time range of 0.1~ms, the peak incident rates presented in Tab.~\ref{tab:PGpeakTubeRates} are acquired.
Due to the distinct energy range and therefore different ToF spectra of the tubes, the highest peak incident rate occurs at different times for each tube of a triplet. The results demonstrate that the peak incident rate on a single detector tube can be as high as 1.7~GHz.

\begin{table}[htp]
\centering
\begin{tabular}{|ccc|}
\hline
Detector tube & McStas [Hz] & Geant4 [Hz] \\\hline
Top & 1.67$\cdot$10$^{9}$ & 1.69$\cdot$10$^{9}$ \\
Central & 1.74$\cdot$10$^{9}$ & 1.75$\cdot$10$^{9}$ \\
Bottom & 1.56$\cdot$10$^{9}$ & 1.56$\cdot$10$^{9}$  \\\hline
\end{tabular}
\vspace{-.1cm}
\caption{\footnotesize Peak incident rate of the 5~meV detector tubes in McStas and Geant4 simulations, with pyrolytic graphite sample.}
\label{tab:PGpeakTubeRates}
\end{table}

\FloatBarrier 
\subsubsection{Yttrium oxide sample}

To give an impression on how the rates change with a different single crystal that is not the highest-case sample but also has a strong Bragg peak, the same simulation and analysis process is repeated using an Y$_2$O$_3$ sample.

\begin{figure}[!h]  
  \centering
  \begin{subfigure}{1.0\textwidth} 
      \includegraphics[width=\textwidth]{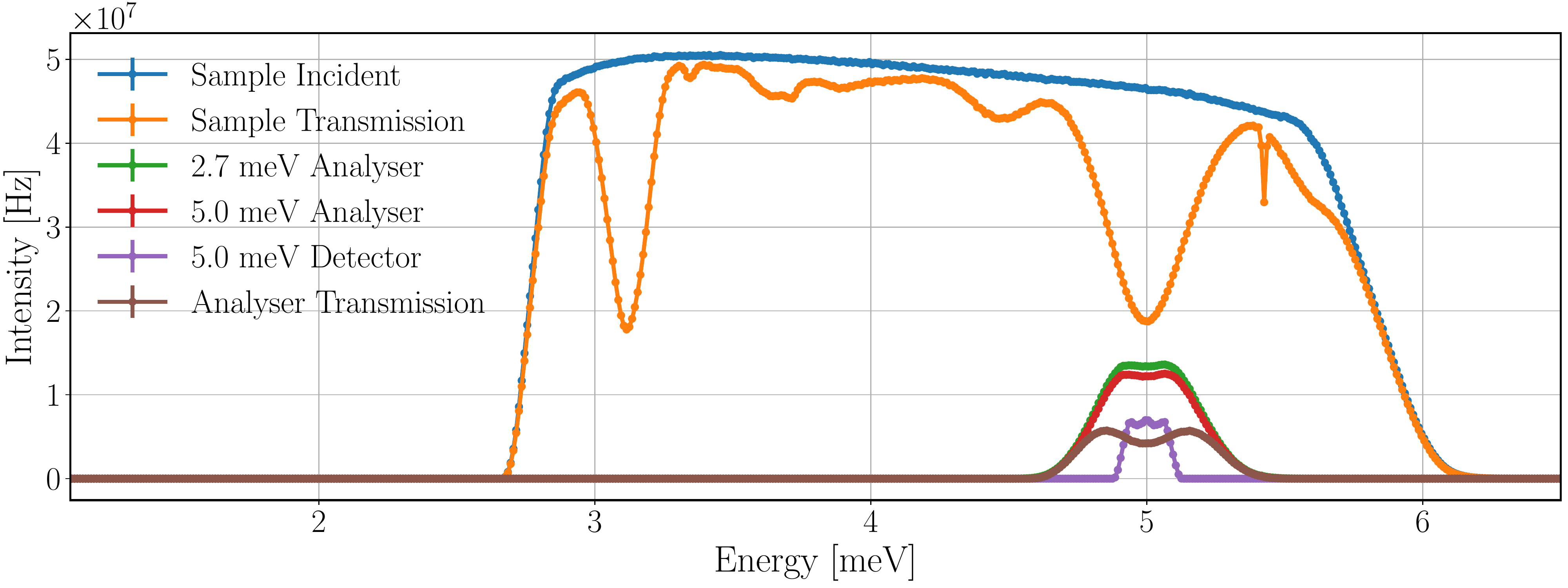} 
     \end{subfigure}
     \begin{subfigure}{1.0\textwidth} 
      \includegraphics[width=\textwidth]{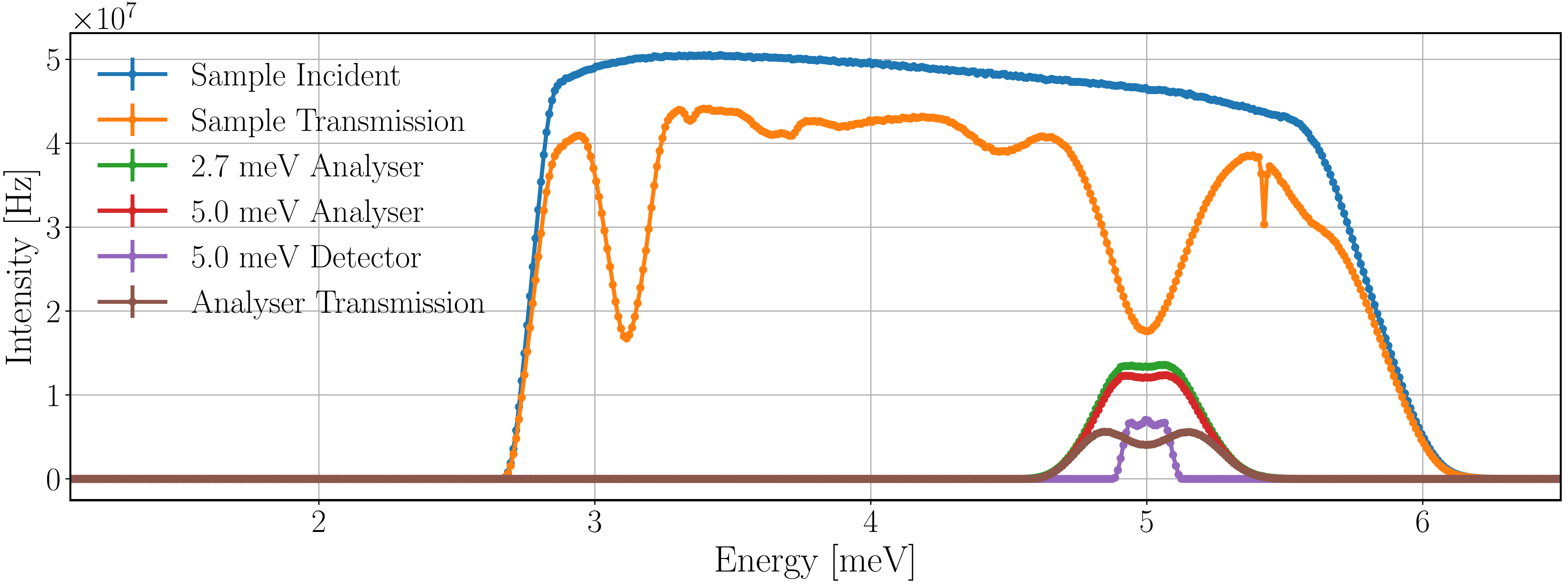}
     \end{subfigure} 
        \caption{\footnotesize Time-averaged neutron energy spectra at the sample and the scattering characterisation system with an Y$_2$O$_3$ single crystal sample in McStas (a) and Geant4 (b) simulation. Incident beam on sample (in blue), beam transmitted through the sample (in orange), beam on the set of analysers for 2.7~meV neutrons (in green), beam on the set of analysers for 5.0~meV neutrons (in red), neutrons hitting the detector triplet for 5.0~meV (in purple), beam transmitted through all sets of analysers (in brown). The lines are only joining the points.}
        \label{fig:changeOfEnergyYO}
\end{figure}

Fig.~\ref{fig:changeOfEnergyYO} demonstrates the change of the time-averaged neutron energy spectrum along the neutrons path at the sample in the scattering characterisation system.
The time-averaged neutron intensities acquired by the integration of the energy spectra are shown in Tab.~\ref{tab:changeOfEnergyYO}.

The results of the McStas and Geant4 simulations agree with only minor differences, with the only exception being the transmission spectrum of the sample.
The transmission of the sample is 10\% higher in McStas, with the same lower intensity bands apparent in the spectrum caused by several crystal planes where the Bragg-criterion is fulfilled for different neutron energies -- including the hkl=-2-2-2 plane for 5.0~meV. 
The source of the discrepancy is the absorption process in the sample, as different absorption cross-sections are used in the two simulation tools, as described earlier in section~\ref{subsec:ncrystal}.

\begin{table}[htp]
\centering
\begin{tabular}{|ccc|}
\hline
Position & McStas [Hz] & Geant4 [Hz]  \\\hline
Sample Incident & 1.45$\cdot$10$^{10}$ & 1.45$\cdot$10$^{10}$ \\
Sample Transmission & 1.24$\cdot$10$^{10}$ & 1.12$\cdot$10$^{10}$ \\
2.7~meV Analyser& 6.01$\cdot$10$^{8}$ & 6.00$\cdot$10$^{8}$ \\
5.0~meV Analyser& 5.52$\cdot$10$^{8}$ & 5.48$\cdot$10$^{8}$ \\
5.0~meV Detector& 1.21$\cdot$10$^{8}$ & 1.21$\cdot$10$^{8}$ \\
Analyser Transmission& 2.96$\cdot$10$^{8}$ & 2.91$\cdot$10$^{8}$ \\\hline
\end{tabular}
\vspace{-.1cm}
\caption{\footnotesize Time-averaged neutron intensities at the sample and different parts of the scattering characterisation system with an Y$_2$O$_3$ single crystal sample in McStas and Geant4 simulation.}
\label{tab:changeOfEnergyYO}
\end{table}

The spectrum of the 2.7~meV analyser arc shows that despite the presence of multiple strong Bragg peaks, it is only the 5.0~meV neutrons that are scattered toward the Q-channel. There are, however, three scattering planes (hkl=2-3-3, hkl=1-2-3 and hkl =1-3-2) on which the 5.0~meV neutrons are Bragg-scattered not toward the Q-channel, causing the slight dip of the peak at 5~meV. 
The spread of neutrons after the sample is less significant than it was with pyrolytic graphite, fewer neutrons are lost on the way toward the 5.0~meV analysers.
 
The time-averaged and peak incident neutron rates for the detector tubes in Tabs.~\ref{tab:YOaverageTubeRates}--\ref{tab:YOpeakTubeRates} show that the neutrons are distributed more evenly among the tubes as a result of the flattened top of the energy spectrum. The maximum of the time-averaged incident rate for a single tube is found to be 41~MHz, with a peak incident rate of a 1~GHz.

\begin{table}[htp]
\centering
\begin{tabular}{|ccc|}
\hline
Detector tube & McStas [Hz] & Geant4 [Hz] \\\hline
Top & 4.08$\cdot$10$^{7}$ & 4.07$\cdot$10$^{7}$ \\
Central & 4.13$\cdot$10$^{7}$ & 4.13$\cdot$10$^{7}$ \\
Bottom & 3.88$\cdot$10$^{7}$ & 3.86$\cdot$10$^{7}$  \\\hline
\end{tabular}
\vspace{-.1cm}
\caption{\footnotesize Time-averaged incident neutron rates of the 5~meV detector tubes in McStas and Geant4 simulations, with an Y$_2$O$_3$ sample.}
\label{tab:YOaverageTubeRates}
\end{table}

\begin{table}[htp]
\centering
\begin{tabular}{|ccc|}
\hline
Detector tube & McStas [Hz] & Geant4 [Hz] \\\hline
Top & 1.03$\cdot$10$^{9}$ & 1.02$\cdot$10$^{9}$ \\
Central & 1.05$\cdot$10$^{9}$ & 1.04$\cdot$10$^{9}$ \\
Bottom & 9.78$\cdot$10$^{8}$ & 9.84$\cdot$10$^{8}$  \\\hline
\end{tabular}
\vspace{-.1cm}
\caption{\footnotesize Peak incident rates of the 5~meV detector tubes in McStas and Geant4 simulations, with an Y$_2$O$_3$ sample.}
\label{tab:YOpeakTubeRates}
\end{table}

Comparing the time-averaged and peak rates to those acquired for pyrolytic graphite, both are lower by a factor of 1.7 but still on the order of 10~MHz for time-averaged and GHz for peak rates.
This means that even with a non-highest-case sample, the rates can be well above the capabilities of the standard $^3$He detector tubes.

As demonstrated in this subsection, the McStas and Geant4 simulation results are in excellent agreement, regarding the detector rates.
For this reason, further simulations are only performed using the Geant4 model of the sample and the scattering characterisation system.
In the subsequent subsections multiple parameters are scanned in order to determine their effect on the incident detector rates, and to prove that the results presented above can be regarded as the highest-case incident rates.
The change in the incident detector rates due to modifying the studied parameters are expected to have the same trend for all single crystals, so all simulations are done using the Y$_2$O$_3$ sample.

\FloatBarrier 
\subsection{Sample mosaicity \label{subsec:sample_mosaicity}}

\begin{figure}[!h]  
  \centering
  \begin{subfigure}{1.0\textwidth} 
      \includegraphics[width=\textwidth]{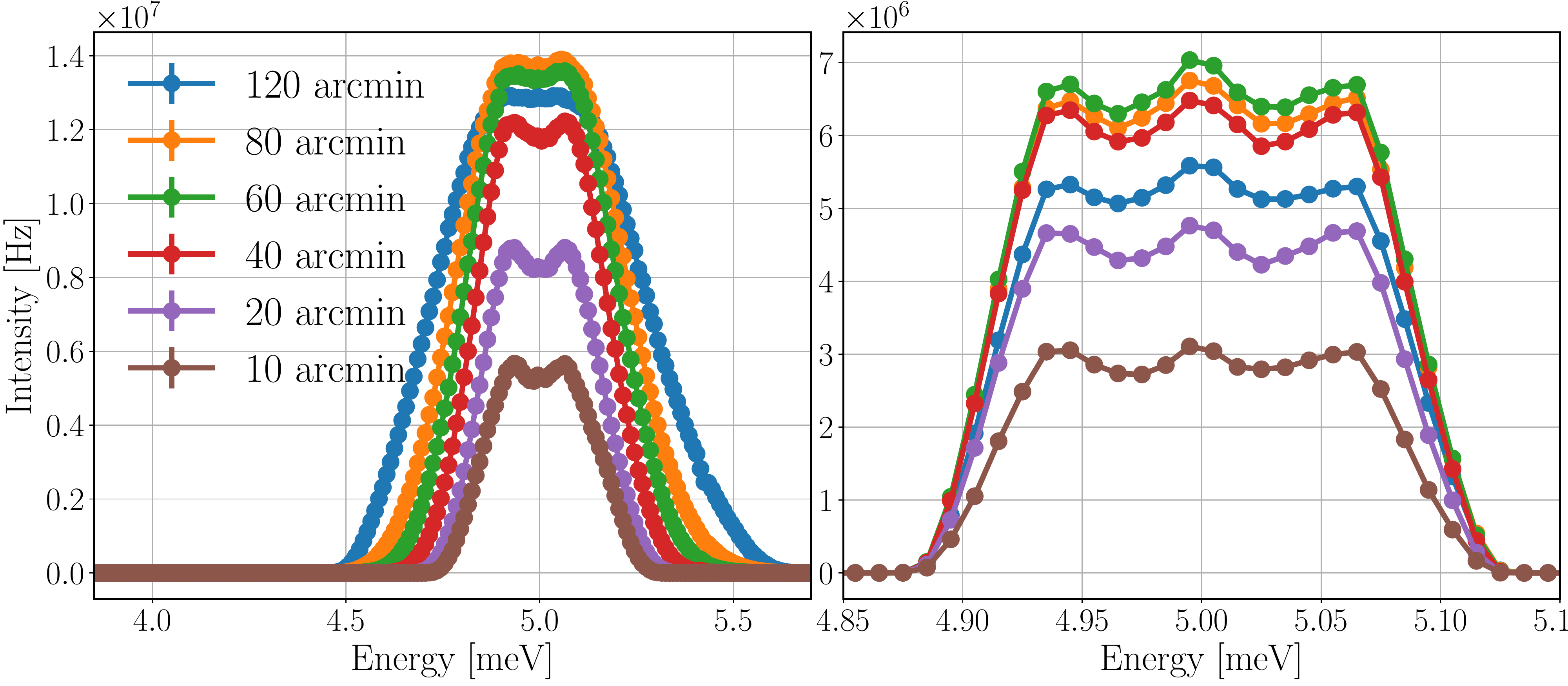}
     \end{subfigure}
   \caption{\footnotesize Energy spectra of the neutron beam on the set of analysers for 2.7~meV neutrons (left) and of the neutrons hitting the detector triplet for 5.0~meV (right) for different sample mosaicities with an Y$_2$O$_3$ sample. The mosaicity of the analysers is 60~arcmin. The lines are only joining the points.}
  \label{fig:sampleMosaicityScan_yo}
\end{figure}

The mosaicity of the sample has multiple effects on the neutron beam Bragg-scattered on a selected scattered plane toward the Q-channel.
A sample with higher mosaicity scatters neutrons of a wider energy range, as the higher spread of crystal plane orientations enables them to fulfil the Bragg-criteria.
This is also true for a wider incident angle range, meaning that neutrons of a divergent beam with higher incident angle have the possibility to be Bragg-scattered on the selected scattering plane.
This higher spread of crystal plane orientations, on the other hand, lowers the probability of neutrons with energy and incident angle close to the ideal values to be scattered. 
The cumulative effect is depicted in Fig.~\ref{fig:sampleMosaicityScan_yo}, showing the energy spectra of the scattered beam at the 2.7~meV analysers and the 5.0~meV detector triplet for an Y$_2$O$_3$ sample with different mosaicities.

The energy spectra of the beam at the 2.7~meV analyser arc shows that the sample with higher mosaicities scatters neutrons of a wider energy range toward the analysers, as expected.
It also shows that the intensities at this point of the instrument are getting higher for mosaicities up until 80~arcmin and therefore intensities for 80~arcmin are higher than those for 60~arcmin.
It is the energy spectrum of neutrons hitting the detector triplet for 5.0~meV, that shows that these additional neutrons do not reach the detectors, as the highest intensities are found at 60~arcmin sample mosaicity.
The resulting time-averaged and peak incident rates of the central detector tube are presented in Tab.~\ref{tab:centralTubeRates_sampleSize}.

\begin{table}[htp]
\centering
\begin{tabular}{|ccc|}
\hline
Mosaicity [arcmin] & Time-averaged rate [Hz] & Peak rate [Hz] \\\hline
120 &  3.30$\cdot$10$^{7}$ &8.37$\cdot$10$^{8}$ \\
80 & 3.99$\cdot$10$^{7}$ & 1.02$\cdot$10$^{9}$ \\
60 & 4.13$\cdot$10$^{7}$ & 1.04$\cdot$10$^{9}$  \\
40 & 3.84$\cdot$10$^{7}$ & 9.63$\cdot$10$^{8}$  \\
20 & 2.80$\cdot$10$^{7}$ & 7.07$\cdot$10$^{8}$  \\
10 & 1.81$\cdot$10$^{7}$ & 4.54$\cdot$10$^{8}$  \\
\hline
\end{tabular}
\vspace{-.1cm}
\caption{\footnotesize Time-averaged and peak incident neutron rates of the central 5~meV detector tube for different sample mosaicities with an Y$_2$O$_3$ sample. The mosaicity of the analysers is 60~arcmin.}
\label{tab:centralTubeRates_sampleSize}
\end{table}

The results are in compliance with the expectation that the highest rates occur when the mosaicity of the sample matches that of the analysers, but also show that within the $\pm$20~arcmin range it is a less than a 10\% effect.

%

\FloatBarrier
\subsection{Analyser mosaicity}

\begin{figure}[!h]  
  \centering
  \begin{subfigure}{1.0\textwidth} 
      \includegraphics[width=\textwidth]{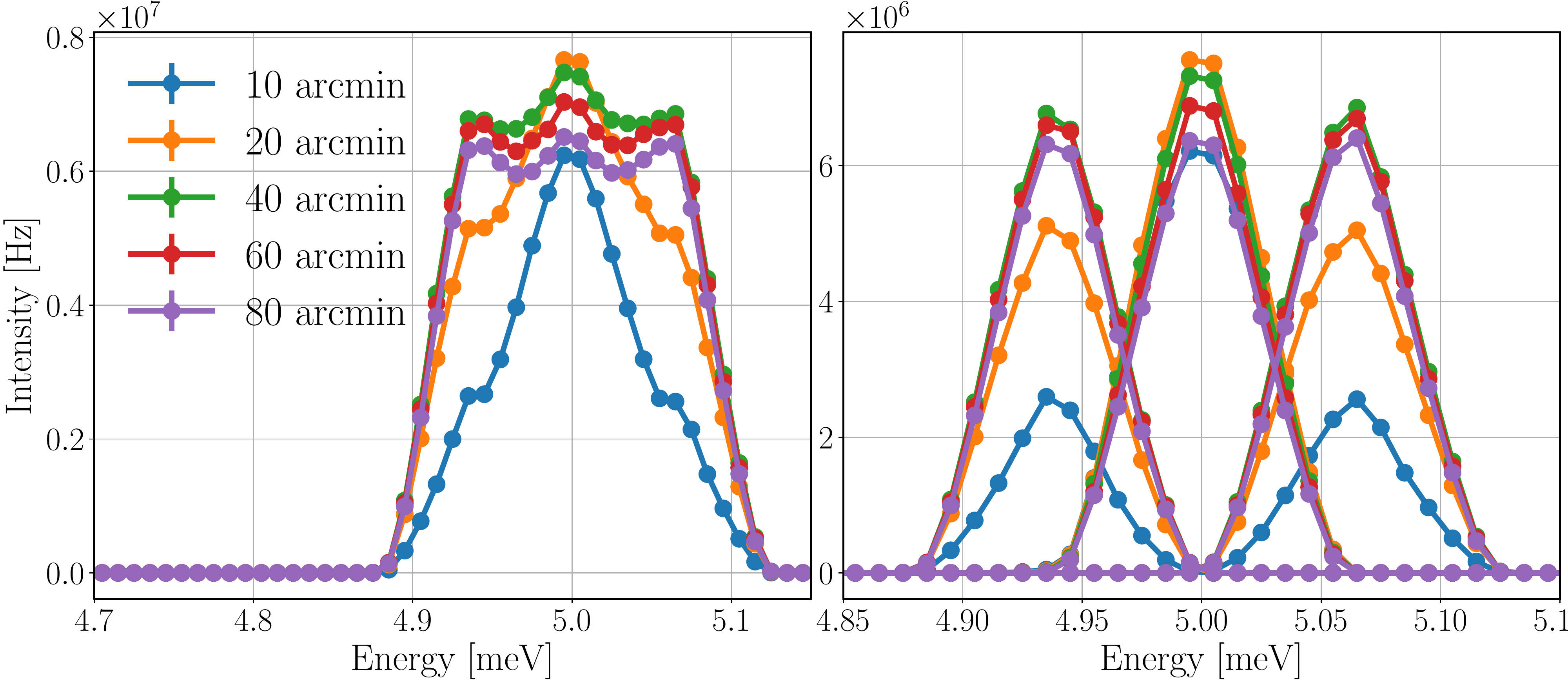}
     \end{subfigure}
   \caption{\footnotesize Energy spectra of the neutrons hitting the detector triplet for the three tubes together (left) and separately (right) for 5.0~meV for different analyser mosaicities and an Y$_2$O$_3$ sample with 60~arcmin mosaicity. The lines are only joining the points.}
  \label{fig:analyserMosaicityScan_yo}
\end{figure}

The mosaicity of the analysers is a fixed value of 60~arcmin for BIFROST, but it is worth briefly investigating how it would affect the rate of the detector tubes. Fig.~\ref{fig:analyserMosaicityScan_yo} depicts the neutron energy spectra of the 5.0~meV detector triplet for different analyser mosaicities with an Y$_2$O$_3$ sample with a mosaicity of 60~arcmin. 

The spectra of the three tubes separately show that for mosaicities below 40~arcmin the two tubes on the sides are under-illuminated compared to the tube in the centre.
In order to apply the prismatic analyser concept, the analyser mosaicity has to be large enough to sufficiently cover all used detectors.
Increasing the mosaicity above 40~arcmin, the intensity in all three tubes is getting slightly lower.
The resulting time-averaged and peak rates of the central detector tube are presented in Tab.~\ref{tab:centralTubeRates_anaMos}.

\begin{table}[htp]
\centering
\begin{tabular}{|ccc|}
\hline
Mosaicity [arcmin] & Time-averaged rate [Hz] & Peak rate [Hz] \\\hline
10 &  4.10$\cdot$10$^{7}$ & 1.04$\cdot$10$^{9}$ \\
20 & 4.68$\cdot$10$^{7}$ & 1.19$\cdot$10$^{9}$ \\
40 & 4.45$\cdot$10$^{7}$ & 1.12$\cdot$10$^{9}$  \\
60 & 4.13$\cdot$10$^{7}$ & 1.04$\cdot$10$^{9}$  \\
80 & 3.84$\cdot$10$^{7}$ & 9.64$\cdot$10$^{8}$  \\
\hline
\end{tabular}
\vspace{-.1cm}
\caption{\footnotesize Time-averaged and peak incident neutron rates of the central 5~meV detector tubes for different analyser mosaicities and an Y$_2$O$_3$ sample with 60~arcmin mosaicity.}
\label{tab:centralTubeRates_anaMos}
\end{table}

The results show that the incident rate in a single detector tube could be 13--14\%~higher in the central tube with 20~arcmin mosaicity compared to the result with 60~arcmin, but the mosaicity has to be higher to apply the prismatic analyser concept, and in the range of 40--80~arcmin the change is less than 10\%.


\FloatBarrier
\subsection{Sample size \label{subsec:sampSize}}

\begin{figure}[!h]  
  \centering
  \begin{subfigure}{0.95\textwidth} 
      \includegraphics[width=\textwidth]{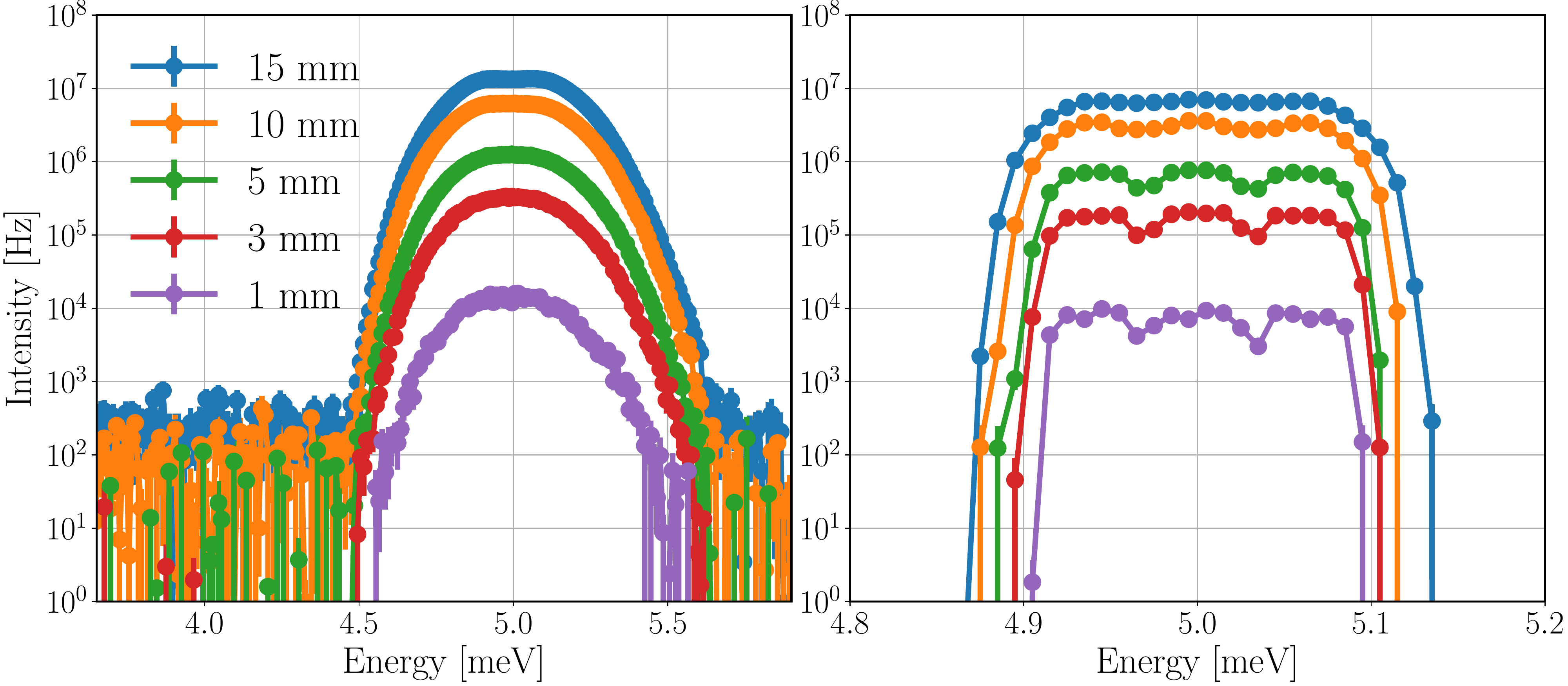}
     \end{subfigure}
   \caption{\footnotesize Energy spectra of the neutron beam on the set of analysers for 2.7~meV neutrons (left) and of the neutrons hitting the detector triplet for 5.0~meV (right) for an Y$_2$O$_3$ sample of different sizes. The diameter and height of the cylindrical samples are equal, with the magnitude indicated in the legend. Both the sample and analyser mosaicity is 60~arcmin. The lines are only joining the points. }
  \label{fig:sampleSizeScan_yo}
\end{figure}

Sample size is the limiting factor in many scientific cases, as it is not easy to grow large samples of some types.
The beam delivery system of BIFROST is optimised for sample cross-sections up to 15$\times$15~mm$^2$ but the realistic sample sizes for the intended applications are much smaller than that, with an expected minimum sample size going down to 1~mm$^3$. 
The heigh, width and thickness of the sample can have different effects on the incident detector rates, however in this parameter scan their cumulative effects are investigate, using cylindrical samples with equal diameter and height. 
The energy spectra of the scattered beam at the 2.7~meV analysers and the 5.0~meV detector triplet for an Y$_2$O$_3$ sample of different sizes are depicted in Fig.~\ref{fig:sampleSizeScan_yo}.
The resulting time-averaged and peak incident rates of the central detector tube are presented in Tab.~\ref{tab:centralTubeRates_sampSize}.

\begin{table}[htp]
\centering
\begin{tabular}{|ccc|}
\hline
Sample size [mm] & Time-averaged rate [Hz] & Peak rate [Hz] \\\hline
15 &  4.13$\cdot$10$^{7}$ & 1.04$\cdot$10$^{9}$  \\
10 & 1.93$\cdot$10$^{7}$ & 4.93$\cdot$10$^{8}$ \\
5 & 3.92$\cdot$10$^{6}$ & 1.01$\cdot$10$^{8}$  \\
3 & 1.04$\cdot$10$^{6}$ & 2.8$\cdot$10$^{7}$  \\
1 & 4.6$\cdot$10$^{4}$ & 1.6$\cdot$10$^{6}$  \\
\hline
\end{tabular}
\vspace{-.1cm}
\caption{\footnotesize Time-averaged and peak incident neutron rates of the central 5~meV detector tube for an Y$_2$O$_3$ sample of different sizes. Both the sample and analyser mosaicity is 60~arcmin.}
\label{tab:centralTubeRates_sampSize}
\end{table}

As expected, the larger the sample, the higher the intensities are. 
By reducing the sample size parameter (height and diameter) from 15~mm to 5~mm and 1~mm, the time-averaged incident rate of the center tube drops by a factor of 10.5 and 900 respectively.
Due to the better resolutions in case of smaller samples the drop in the peak incident rate is lower, a factor of 10.3 for 5~mm and a factor of 650 for 1~mm.

Another effect of the better resolution is visible in the energy spectra of the detector triplets, where the three-peak structure is more apparent for smaller samples.

\FloatBarrier
\subsection{Pulse-shaping chopper opening time \label{subsec:psc}}

The energy resolution of the instrument can be increased at the cost of neutron intensity by modifying the opening time of the pulse-shaping chopper. 
The flux on the sample and the detectors are both expected to drop significantly in case of the high resolution setting, when the opening time is merely 0.1~ms, compared to the high flux mode achieved by an opening time of 5~ms.

\begin{figure}[!h]  
 \centering
 \includegraphics[width=\textwidth]{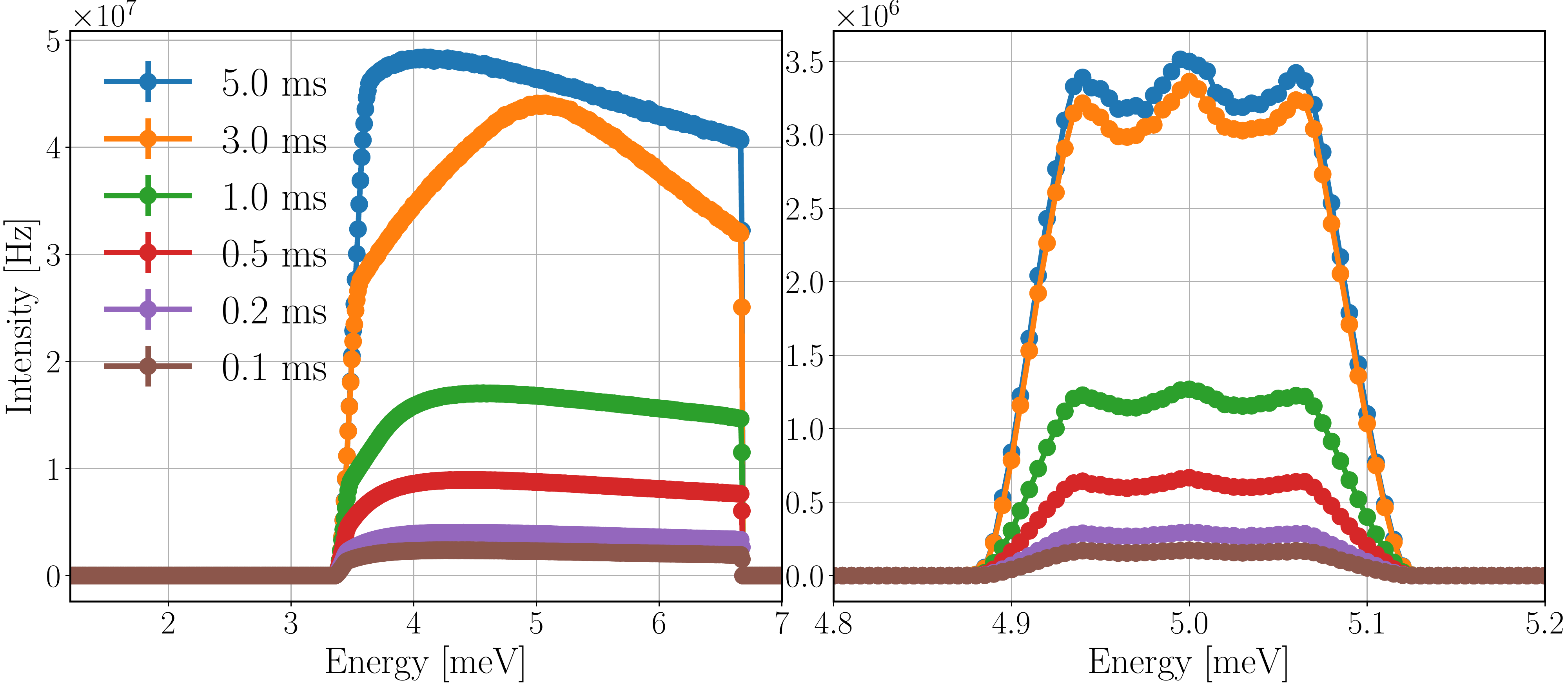}
 \caption{\footnotesize Energy spectra of the neutron beam on the sample (left) and of the neutrons hitting the detector triplet for 5.0~meV (right) for an Y$_2$O$_3$ sample for different pulse-shaping chopper opening times. The lines are only joining the points.}
 \vspace{-.1cm}
 \label{fig:pscScan_yo}
\end{figure}

The energy spectra of the neutron beam at the sample and the 5.0~meV detector triplet for an Y$_2$O$_3$ sample for different pulse-shaping chopper opening times are depicted in Fig.~\ref{fig:pscScan_yo}, with the time-averaged intensities presented in Tab.~\ref{tab:pscRates}.
The time-averaged and peak incident rates of the central  5.0~meV detector tube are presented in Tab.~\ref{tab:centralTubeRates_psc}.

\begin{table}[!h] 
\begin{center}
\begin{tabular}{|ccc|}
\hline
PSC opening time [ms] & Sample [Hz] & 5.0~meV detectors [Hz] \\\hline
5.0 & 1.44$\cdot$10$^{10}$ & 1.21$\cdot$10$^{8}$ \\
3.0 & 1.22$\cdot$10$^{10}$ &  1.14$\cdot$10$^{8}$ \\
1.0 & 5.05$\cdot$10$^{9}$ & 4.38$\cdot$10$^{7}$ \\
0.5 & 2.67$\cdot$10$^{9}$ & 2.27$\cdot$10$^{7}$ \\
0.2 & 1.20$\cdot$10$^{9}$ & 1.01$\cdot$10$^{7}$ \\
0.1 & 7.03$\cdot$10$^{8}$ & 5.93$\cdot$10$^{6}$  \\\hline
\end{tabular}
\vspace{-.1cm}
\caption{\footnotesize Time-averaged neutron intensities at the sample and the 5.0~meV detector tubes with an Y$_2$O$_3$ sample for different pulse-shaping chopper (PSC) opening times.}
\label{tab:pscRates}
\end{center}
\end{table}

\begin{table}[!h] 
\begin{center}
\begin{tabular}{|ccc|}
\hline
PSC opening time [ms]  & Time-averaged rate [Hz] & Peak rate [Hz] \\\hline
5.0 & 4.13$\cdot$10$^{7}$ & 1.04$\cdot$10$^{9}$  \\
3.0 & 3.91$\cdot$10$^{7}$ & 1.04$\cdot$10$^{9}$ \\
1.0 & 1.50$\cdot$10$^{7}$ & 8.76$\cdot$10$^{8}$  \\
0.5 & 7.79$\cdot$10$^{6}$ & 5.31$\cdot$10$^{8}$  \\
0.2 & 3.46$\cdot$10$^{6}$ & 2.52$\cdot$10$^{8}$  \\
0.1 & 2.03$\cdot$10$^{6}$ & 1.50$\cdot$10$^{8}$  \\
\hline
\end{tabular}
\vspace{-.1cm}
\caption{\footnotesize Time-averaged and peak incident neutron rates of the central 5~meV detector tube with Y$_2$O$_3$ sample for different pulse-shaping chopper opening times.}
\label{tab:centralTubeRates_psc}
\end{center}
\end{table}

As expected, the time-averaged rates on the sample and on the detectors decrease with shorter pulse-shaping chopper opening times. The difference in time-averaged incident rates between the high flux mode (5~ms) and the high resolution mode (0.1~ms) is approximately a factor of 20 for both the detector triplet and the central tube.

\begin{figure}[!h]  
 \centering
 \includegraphics[width=\textwidth]{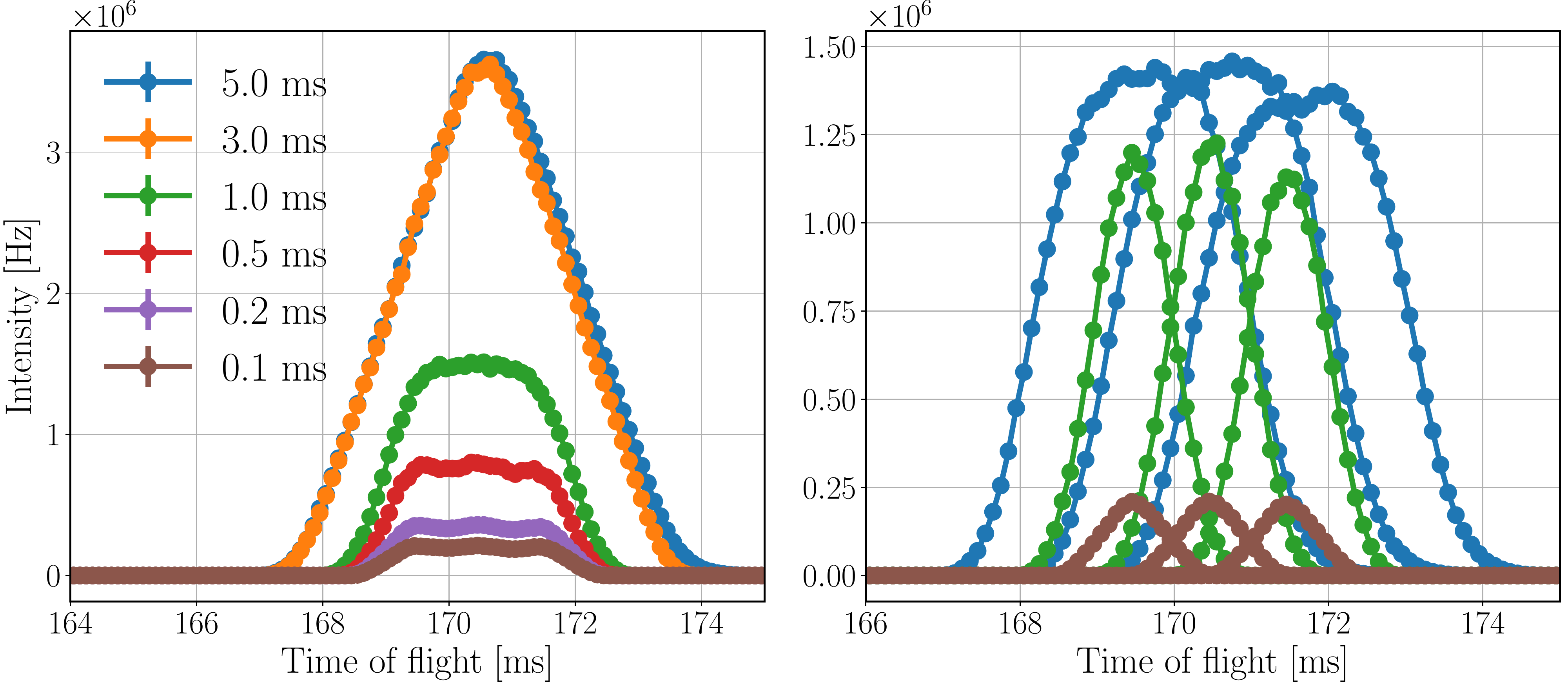}
  \caption{\footnotesize ToF spectrum of neutrons at the 5~meV detector triplet for the three tubes together (on left) and separately (on right) with Y$_2$O$_3$ sample for different pulse-shaping chopper opening times. 
   The separated spectrum is not displayed in all cases in the righthand figure to avoid the figure being overcrowded.
   The lines are only joining the points.}
  \label{fig:detectorTofPsc_yo}
\end{figure}

Regarding the peak rates in the central tube, however, this drop is less apparent. 
The highest rate for a 3~ms opening time is the same (within statistical uncertainty) as the rate for 5~ms, and the difference compared to 0.1~ms opening time is only a factor of 6.8. 
The reason for this difference is the better ToF resolution with shorter pulse-shaping chopper opening times.
The higher time-averaged rates are distributed over a longer period of time on the detectors, as it is demonstrated in Fig.~\ref{fig:detectorTofPsc_yo}, showing the effect of the pulse-shaping chopper opening time on the ToF spectrum of neutrons hitting the detectors.


The longer the opening time, the broader the ToF peaks are. This directly affects the energy resolution, and increases the dead time in case of saturation.
If one tube of a triplet is saturated, then none of the three can read out data, as they are connected in series. This means that for the detector triplet in the presented case for 5.0~ms opening time no data is recorded for more than 6~ms.

\FloatBarrier 
\section{Elastic peak rates in representative operational conditions}

Parameters chosen in section~\ref{mcstasGeantComparison} correspond to possible highest-case scenarios and rates acquired are far above the capabilities of $^3$He tubes. 
However, the combination of a strongly scattering large sample and the highest flux mode is rather artificial, so it is worth evaluating a more representative operational scenario. 

BIFROST is designed for small samples, as sample size is the limiting factor in many science cases. Hence, cm-size crystals are not to be expected very often, only large samples with small magnetic moments, and therefore small magnetic Bragg peak intensity. 
There is another parameter directly affecting the intensities but not discussed yet, the accelerator power of the ESS source. 
As mentioned earlier in section~\ref{subsec:mcstas}, the source power of 5~MW is used for the simulations.
That is the eventual operational power of ESS, however, it will initially operate at 2~MW.
The intensities are expected to scale linearly with the source power. 

\begin{figure}[!h]  
  \centering
  \begin{subfigure}{1.0\textwidth} 
      \includegraphics[width=\textwidth]{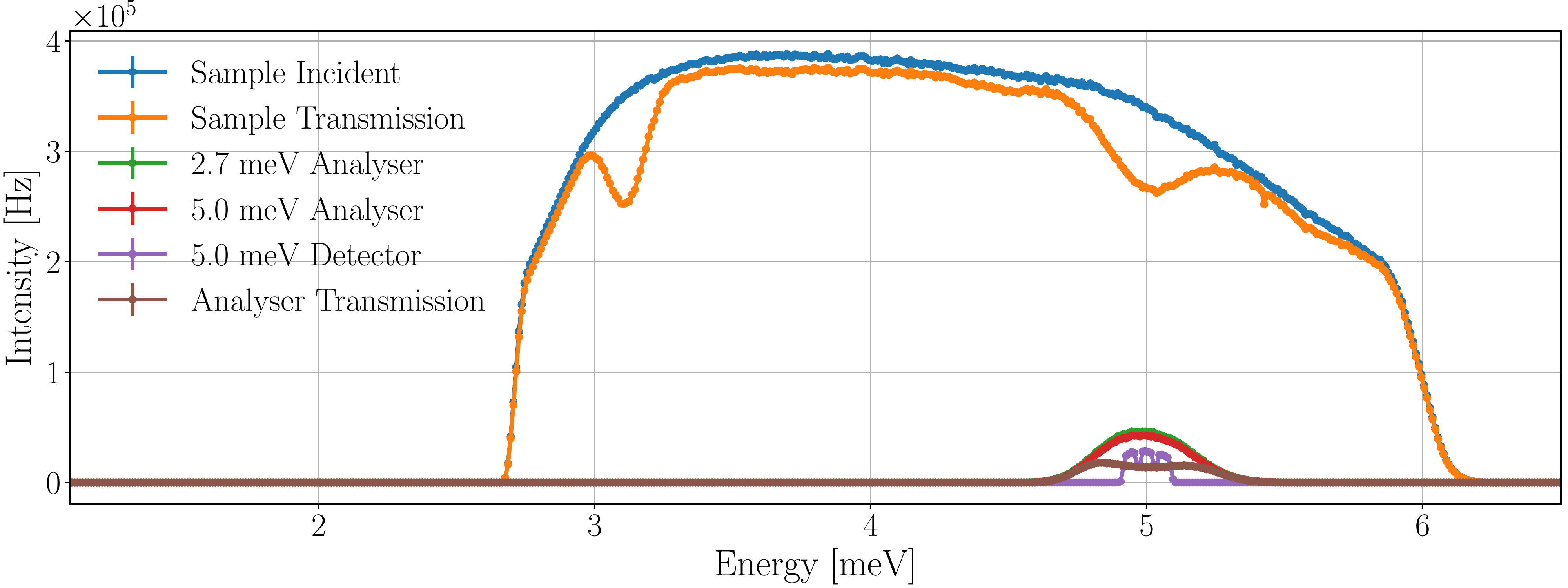}
     \end{subfigure}
   \caption{\footnotesize Time-averaged neutron energy spectra at the sample and the scattering characterisation system with an Y$_2$O$_3$ sample in Geant4 simulation using 2~MW source power, 3~mm height and diameter sample size and 60~arcmin sample mosaicity. Incident beam on sample (in blue), beam transmitted through the sample (in orange), beam on the set of analysers for 2.7~meV neutrons (in green), beam on the set of analysers for 5.0~meV neutrons (in red), neutrons hitting the detector triplet for 5.0~meV (in purple), beam transmitted through all sets of analysers (in brown). The lines are only joining the points.}
   \label{fig:changeOfEnergyYO_2MW}
\end{figure}

For these reasons, the following parameters are selected to define the rates in a more representative operational case:
2~MW source power, 1~ms pulse-shaping chopper opening time, an Y$_2$O$_3$ single crystal sample with a height and diameter of 3~mm, and mosaicity of 60~arcmin. 
The time-averaged energy spectra of the neutron beam at the sample and different positions of the scattering characterisation system, acquired with these parameters are demonstrated in Fig.~\ref{fig:changeOfEnergyYO_2MW}, along with the integral values presented in Tab.~\ref{tab:changeOfEnergyYO_2MW}.

\begin{table}[h!]
\begin{center}
\begin{tabular}{|cc|}
\hline
Position & Intensity [Hz]  \\\hline
Sample Incident & 1.07$\cdot$10$^{8}$ \\
Sample Transmission & 1.01$\cdot$10$^{8}$ \\
2.7~meV Analyser& 1.84$\cdot$10$^{6}$  \\
5.0~meV Analyser& 1.70$\cdot$10$^{6}$ \\
5.0~meV Detector& 4.1$\cdot$10$^{5}$  \\
Analyser Transmission& 8.71$\cdot$10$^{5}$  \\\hline
\end{tabular}
\vspace{-.1cm}
\caption{\footnotesize Time-averaged neutron intensities at the sample and different parts of the scattering characterisation system with an Y$_2$O$_3$ sample, using 2~MW source power, 1~ms PSC opening time, 3~mm height and diameter sample size and 60~arcmin sample mosaicity.}
\label{tab:changeOfEnergyYO_2MW}
\end{center}
\end{table}

The combined effect of the lower source power, shorter pulse-shaping chopper opening time and smaller sample (cross-section) decreased the time-averaged neutron intensity on the sample significantly, by a factor of 136 compared to the highest-case scenario with an Y$_2$O$_3$ sample.
Due to the reduced sample thickness, the transmission through the sample is increased to 94\% from 77\%, as a result of lower absorption and weaker Bragg peaks.
The lower incident intensity on the sample, and weaker Bragg peak lead to a drop by a factor of 322--326 in the time-averaged neutron intensity on both the 2.7 and 5.0~meV analysers.
The drop in the time-averaged neutron intensity on the 5.0~meV detector triplet is a little bit lower, a factor of 295, due to the smaller divergence and better energy resolution of the neutron beam compared to the highest-case scenario.

The resulting time-averaged and peak incident neutron rates of the 5~meV detector tubes are presented in Tab.~\ref{tab:YOTubeRates_2MW}.
The highest time-averaged incident neutron rate on a single tube is found to be 0.15~MHz, that means a drop by a factor of 275, but the peak incident rate is 9.9~MHz, which is lower only by a factor 105 compared to the highest-case results.

\begin{table}[htp]
\begin{center}
\begin{tabular}{|ccc|}
\hline
Detector tube & Time-averaged rate [Hz] & Peak rate [Hz] \\\hline
Top &  1.3$\cdot$10$^{5}$ & 9.3$\cdot$10$^{6}$ \\
Central & 1.5$\cdot$10$^{5}$ & 9.9$\cdot$10$^{6}$ \\
Bottom & 1.3$\cdot$10$^{5}$ & 9.4$\cdot$10$^{6}$  \\\hline
\end{tabular}
\vspace{-.1cm}
\caption{\footnotesize Time-averaged and peak incident neutron rates of the 5~meV detector tubes with an Y$_2$O$_3$ sample, using 2~MW source power, 1~ms PSC opening time, 3~mm height and diameter sample size and 60~arcmin sample mosaicity.}
\label{tab:YOTubeRates_2MW}
\end{center}
\end{table}

The numbers and reduction factors are in accordance with previous simulations in sections~\ref{subsec:sampSize} and \ref{subsec:psc} where the effect of sample size and pulse-shaping chopper opening time were investigated separately. 

There are factors not taken into account in the current study, that may further reduce the intensities slightly on the detectors, like the non-ideal transmission of the filtering system, and the effect of the divergence jaws applicable for reducing the angular spread of neutrons.

\FloatBarrier 
\section{Simulation with full scattering characterisation system \label{section:vanadium}}

The previous sections were aimed to define the highest incident rates a detector tube can experience using different instrument and sample parameters in case of a coherent elastic (Bragg) peak. 
This section intends to present the incident detector rates in case of incoherent elastic peaks with a standard calibration sample, and to demonstrate the use of the full simulation model of the scattering characterisation system with all nine Q-channels.

\begin{figure}[!h]  
  \centering
       \begin{subfigure}{0.7\textwidth} 
      \includegraphics[width=\textwidth]{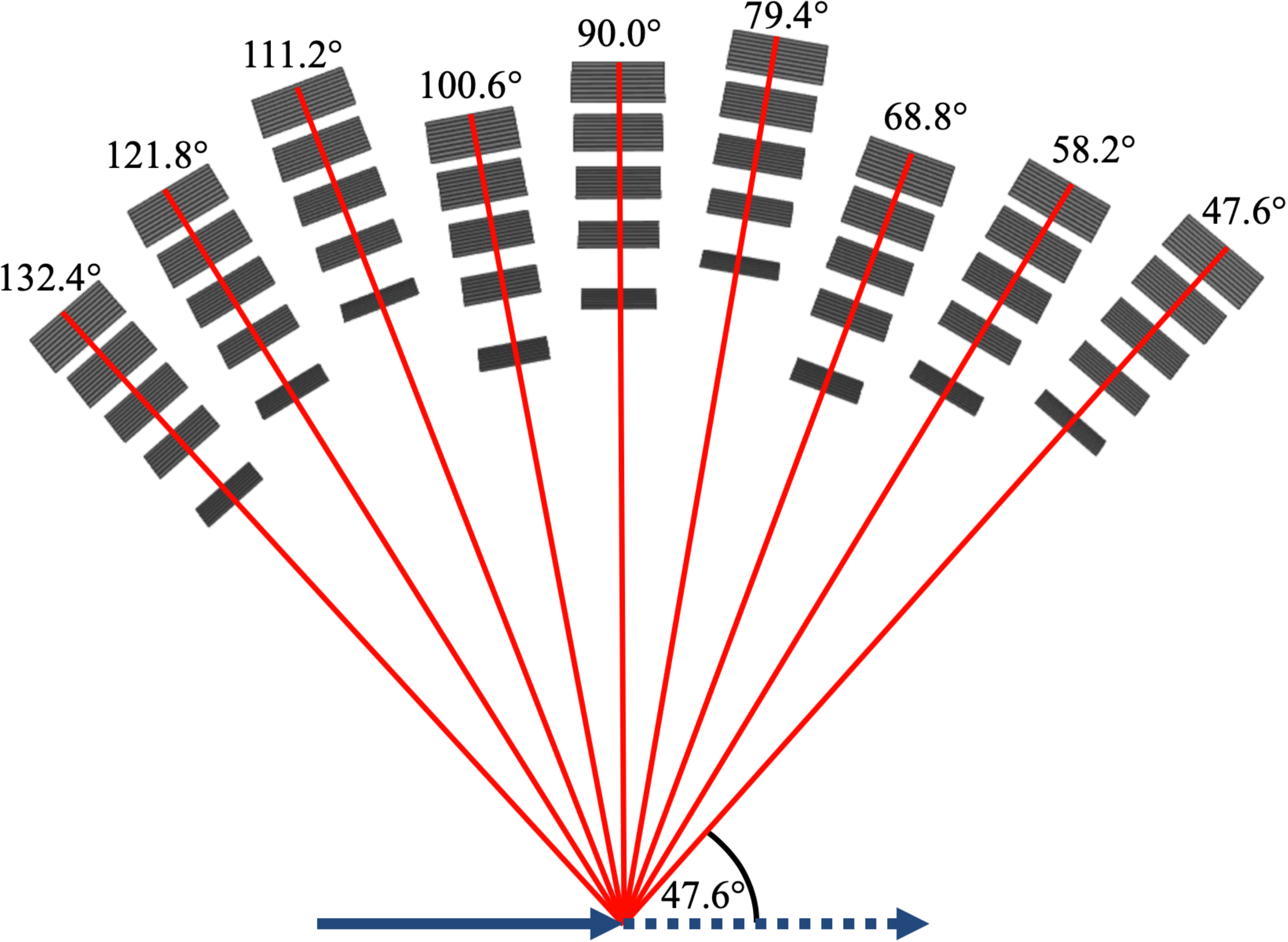}
     \end{subfigure}
   \caption{\footnotesize Top view schematic figure of the scattering characterisation system model with all nine Q-channels. The red lines and corresponding angles indicate the scattering angle for the centre of each Q-channel.}
   \label{fig:Qchannels}
\end{figure}

The sample selected for this simulation is vanadium, that is assumed to be an incoherent elastic scatterer which scatters isotropically and therefore it is used to calibrate the incident neutron intensity and the detector efficiencies in neutron spectrometers~\cite{MAYERS1984609}.
As mentioned earlier, in previous simulations the single Q-channel present in the model was rotated according to the Bragg-angle of the sample for 5~meV neutrons ($\Theta=37.067^\circ$ for pyrolytic graphite and $\Theta=41.169^\circ$ for Y$_2$O$_3$). For vanadium the rotation of the nine Q-channels is arbitrarily selected in a way to have $2\Theta=90^\circ$ scattering angle for the central Q-channel, as depicted in Fig.~\ref{fig:Qchannels}.

\begin{figure}[!h]  
  \centering
       \begin{subfigure}{1.0\textwidth} 
      \includegraphics[width=\textwidth]{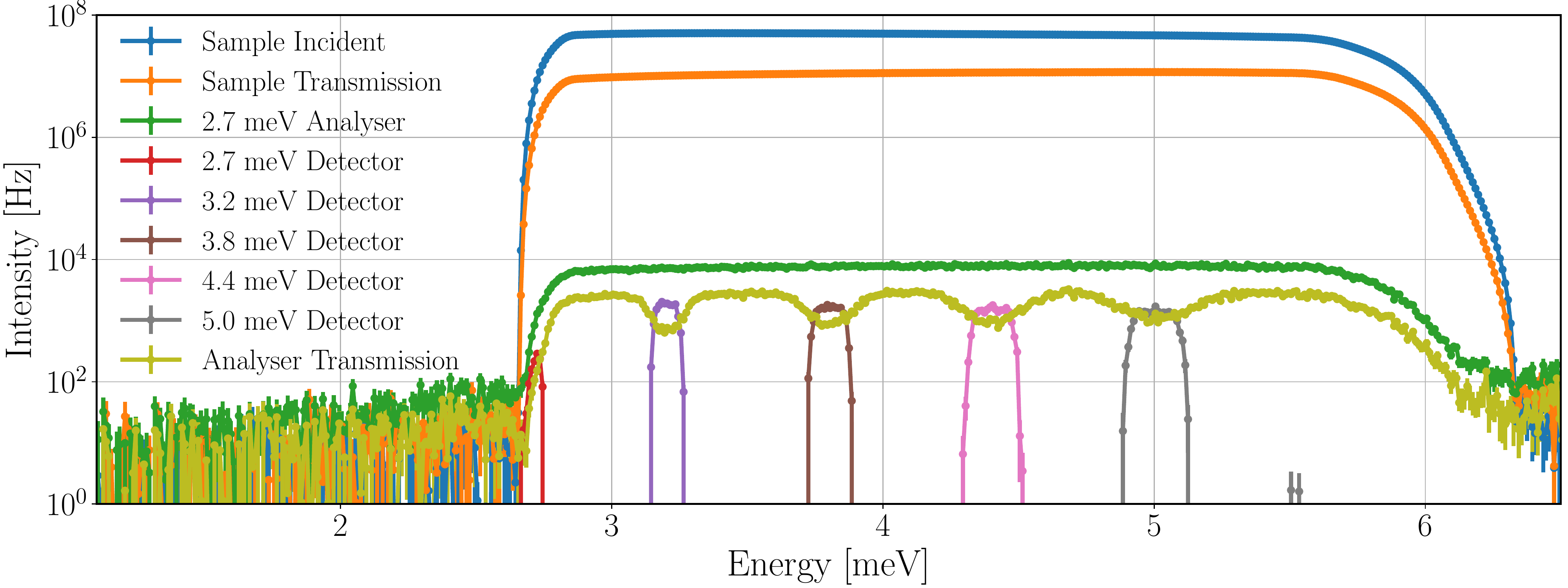}
     \end{subfigure}
    \caption{\footnotesize Time-averaged neutron energy spectra at the sample and in the central Q-channel with a vanadium sample in Geant4 simulation. Incident beam on sample (in blue), beam transmitted through the sample (in orange), beam on the set of analysers for 2.7~meV neutrons (in green), neutrons hitting the detector triplets for energies 2.7--5.0~meV (in red--grey), beam transmitted through all sets of analysers (in mustard). The lines are only joining the points.}
   \label{fig:changeOfEnergyV}
\end{figure}

The instrument and sample parameters are the same as for the highest-case scenario: 5~MW source power, 5~ms pulse-shaping chopper opening time, 15~mm sample height and diameter. 
The time-averaged energy spectra of the neutron beam at the sample and different positions of the central Q-channel are demonstrated in Fig.~\ref{fig:changeOfEnergyV}, along with the integral values presented in Tab.~\ref{tab:changeOfEnergyVanadium}.

\begin{table}[htp]
\begin{center}
\begin{tabular}{|cc|}
\hline
Position & Intensity [Hz]  \\\hline
Sample Incident & 1.45$\cdot$10$^{10}$ \\
Sample Transmission & 3.36$\cdot$10$^{9}$ \\
2.7~meV Analyser& 2.55$\cdot$10$^{6}$  \\
2.7~meV Detector& 1.6$\pm 0.3\cdot$10$^{3}$  \\
3.2~meV Detector& 1.7$\pm 0.1\cdot$10$^{4}$  \\
3.8~meV Detector& 2.1$\pm 0.1\cdot$10$^{4}$  \\
4.4~meV Detector& 2.4$\pm 0.1\cdot$10$^{4}$  \\
5.0~meV Detector& 2.6$\pm 0.2\cdot$10$^{4}$  \\
Analyser Transmission& 7.20$\cdot$10$^{5}$  \\\hline
\end{tabular}
\vspace{-.1cm}
\caption{\footnotesize Time-averaged neutron intensities at the sample and different parts of the scattering characterisation system with a vanadium sample in Geant4 simulation. }
\label{tab:changeOfEnergyVanadium}
\end{center}
\end{table}

The transmission through the sample is only 23\% -- much lower than it is for pyrolytic graphite (80\%) or Y$_2$O$_3$ (77\%) -- with no peaks missing from the spectrum, as expected from a sample scattering mainly incoherently. 
The wide spectrum of the 2.7~meV analysers also shows that neutrons are not coming from an elastic peak, but despite the wider energy range, a logarithmic scale is needed as the integrated intensity is more than 2 orders of magnitude lower than experienced with previous samples.
The spectrum of the beam transmitted through all five sets of analysers clearly shows neutrons missing because they are selected by the analysers.
These neutrons appear in the spectra of the detectors, that show that peaks are narrower for lower energies as a result of the better energy resolution of the analysers for lower energies.
The incident intensity on the 2.7~meV detectors is much lower than in other detectors due the energy range selected by the bandwidth chopper.
The resulting time-averaged incident rates are higher for higher energies, with the maximum of 26$\pm$2~kHz for the 5.0~meV detector triplet.
For a single detector tube the highest rates are found for the central 5.0~meV detector with a time-averaged intensity of 9$\pm$1~kHz and peak intensity of 0.3$\pm$0.1~MHz.
The time-averaged incident neutron rates of all detector triplets in all Q-channels are presented in Tab.~\ref{tab:fullGeomVanadium}.

\begin{table}[htp]
\begin{center}
\begin{tabular}{|ccccccc|}
\hline
Q-& Scattering & 2.7~meV & 3.2~meV & 3.8~meV & 4.4~meV  & 5.0~meV \vspace{-0.15cm}\\ 
channel&angle [$^\circ$] &[kHz] &[kHz]  &[kHz]  & [kHz]  & [kHz] \\
\hline
1&132.4 & 2.8$\pm 0.3$& 23$\pm 1$& 28$\pm 1$& 30$\pm 2$& 31$\pm 2$\\
2&121.8 & 2.2$\pm 0.3$& 20$\pm 1$& 23$\pm 1$& 27$\pm 2$& 30$\pm 2$\\
3&111.2& 1.9$\pm 0.3$& 18$\pm 1$& 21$\pm 1$& 24$\pm 1$& 25$\pm 1$\\\hline
4&100.6 & 2.3$\pm 0.3$& 21$\pm 1$& 25$\pm 1$& 28$\pm 2$& 30$\pm 2$\\
5&90.0& 1.6$\pm 0.3$& 17$\pm 1$& 21$\pm 1$& 24$\pm 1$& 26$\pm 2$\\
6&79.4 & 1.4$\pm 0.2$& 15$\pm 1$& 18$\pm 1$& 21$\pm 1$& 22$\pm 2$\\\hline
7&68.8 & 2.0$\pm 0.2$& 18$\pm 1$& 22$\pm 1$& 25$\pm 2$& 26$\pm 1$\\
8&58.2 & 1.4$\pm 0.2$& 15$\pm 1$& 19$\pm 1$& 22$\pm 1$& 23$\pm 1$\\
9&47.6 & 1.1$\pm 0.2$& 13$\pm 1$& 16$\pm 1$& 19$\pm 1$& 21$\pm 1$\\\hline
\end{tabular}
\vspace{-.1cm}
\caption{\footnotesize Time-averaged neutron intensities of the 5 detector triplets in all nine Q-channels with a vanadium sample.}
\label{tab:fullGeomVanadium}
\end{center}
\end{table}

The trends in the results demonstrate the combination of three effects.
In each Q-channel the detector triplets for higher energies experience higher incident rates due to the wider energy ranges selected by the analysers, as shown for the central Q-channel earlier in this section.
The second effect has roots in the ``triple stagger'' geometry and the asymmetry of the Q-channels described in section~\ref{subsec:instrument}. 
The sample--analyser distances in Q-channels 1, 4, 7 are shorter, and in channels 3, 6, 9 are longer than distances in channels 2, 5, 8.
The shorter distances increase the rates visibly because neutrons are not collimated by Bragg-scattering on the sample and therefore their spread at longer distances becomes important. 
This effect on the rates is somewhat blurred by the third effect caused by the anisotropy of the scattering cross-section in vanadium.
For the observed energies the scattering cross-section of vanadium is slightly higher for higher scattering angles \cite{MAYERS1989654}, and more importantly, the neutron path length through the solid cylindrical sample is generally higher for neutrons scattered in lower angles, therefore the absorption is higher for these neutrons.
These two effects result in generally higher rates for Q-channels positioned for higher scattering angles, but due to the asymmetry of the adjacent Q-channels, it is most apparent when comparing Q-channels of the same symmetry, like 2, 5, 8.

\FloatBarrier
\section{Conclusions}

BIFROST is an indirect ToF spectrometer at ESS and one of the first eight instruments to be constructed.
One of the most challenging aspects of its operation is the rate capability and in particular the peak instantaneous rate capability, i.e.\,the number of neutrons hitting the detector per channel at the peak of the neutron pulse. 
There is no intent to measure the intensity of elastic peaks as they are considered background for this instrument, however it is vital that the detectors are not degraded by such intensity and remain capable of measuring weak inelastic signals, as soon as possible after saturation.
This implies that the detector aspects of recovery time and high rate tolerance have to be carefully evaluated by measurements to prove that scientific performance will be intact.

A detailed methodology for acquiring the results is presented.
The full simulation of the instrument from source to detector position is carried out using multiple simulation software packages.
A flexible model of the sample and the scattering characterisation system of BIFROST is implemented in both McStas and Geant4 and a comparison of their strengths and weaknesses is presented. 
The capability of both simulation tools is enhanced by the NCrystal library and associated tools. 
The first application of the special NCrystal pyrolytic graphite is presented, demonstrating its capabilities for modelling analysers for neutron scattering applications. 

McStas is capable of simulating instruments as long as 160~m, and even handling beam splitting to some extent, to treat simulations with multiple set of analysers, however the latter comes with great complexity and some limitations, as it is not within the natural usage of this simulation software. 
Geant4 on the other hand is not suited to simulate the beam transport system of an instrument, but with the use of NCrystal, it is an entirely appropriate tool for a scattering characterisation system with any level of geometrical complexity, and even offers the possibility to include parts like filtering system and cross-talk shielding, and taking into account back-scattering. 
The results of the McStas and Geant4 model of the scattering characterisation system are compared using various single crystal samples.
The results show perfect agreement with the only exception being the transmission through the sample where a difference less than 10\% is found in one case, due to the more detailed modelling of absorption in Geant4.

With this knowledge at hand a choice was made to combine the McStas model of the beam transport system and the Geant4 model of the sample and the scattering characterisation system using the MCPL tools. 
Using this model, the incident detector rates anticipated at the BIFROST instrument for different configurations are presented.
The impact of sample type, sample and analyser mosaicity, sample size, and pulse shaping chopper opening time is studied on the incident detector rates. 
For instrument configurations and sample parameters representing highest-case conditions, it is determined that the peak rate can reach the value of 1--1.7~GHz for a single detector tube with time-averaged rates of 40--70~MHz. 
These tubes are expected to reach saturation well below that, at 50--100~kHz. 
These tubes will also be saturated for a minimum of 5~ms, but the saturation deadtime for detecting signals is more like 6~ms because the counting detector tubes are coupled in triplets.

To overcome challenges caused by these rates, an operational evaluation of a measurement strategy will be the key to the successful operation of this instrument. 
More ``everyday'' realistic samples give a lower rate challenge, however these samples will still saturate detectors. 

A simulation with the full analyser system is presented using a common calibration sample. 
This model can now be used to predict experimental conditions from specific proposed samples, i.e.\,sample size and composition for experiment planning purposes for users.

The results here show the potential power of source to detector simulation for neutron scattering. These simulations are now possible due to tools recently developed. It is now possible to realistically simulate very complex systems. 

\FloatBarrier
\section*{Acknowledgements}

This work has been supported by the In-Kind collaboration between
ESS ERIC (contract number: NIK5.4 \#10 [ESS]) and the Hungarian
Academy of Sciences, Centre for Energy Research (MTA EK).
The authors would like to thank the DMSC Computing Centre which provided the computing resources for the simulations~\cite{dmscURL}.
Richard Hall-Wilton, Kalliopi Kanaki and Thomas Kittelmann would like to acknowledge 
support from BrightnESS [EU Horizon2020 grant 676548].

\FloatBarrier

\FloatBarrier


\bibliography{refs_mk.bib}

\end{document}